\documentclass[12pt]{article}

\usepackage[utf8]{inputenc}
\usepackage{amsmath, amssymb}
\usepackage{graphicx}
\usepackage{natbib}
\usepackage{geometry}
\usepackage{setspace}
\usepackage{booktabs}
\usepackage{hyperref}
\usepackage{caption}
\usepackage{subcaption}
\usepackage{xcolor}
\usepackage{soul}

\soulregister\citep7
\soulregister\ref7
\soulregister\texttt7

\title{A Bayesian Framework for Extrapolative Emulation of Spatially Gridded Simulation Data}
\author{
Kelly R. Moran\textsuperscript{1,*}, Ky Potter\textsuperscript{1}, Chris Danly\textsuperscript{2}, Christopher L. Fryer\textsuperscript{3} \\
\\
\textsuperscript{1}\small Statistical Sciences Group, Los Alamos National Laboratory \\
\textsuperscript{2}\small Dynamic Imaging and Radiography Group, Los Alamos National Laboratory \\
\textsuperscript{3}\small Center for Theoretical Astrophysics, Los Alamos National Laboratory \\
\textsuperscript{*}\small Corresponding author: \texttt{krmoran@lanl.gov}
}
\date{\today}

\begin{document}

\maketitle

\begin{abstract}
We propose a Bayesian emulator for extrapolating spatially gridded simulation output across resolution. The method treats each pixel as following a nonlinear resolution-response curve, while linking pixels through Gaussian process priors on the curve parameters to preserve spatial structure and quantify uncertainty. In synthetic experiments designed to mimic resolution-dependent bias, the approach recovers the high-resolution target with competitive or improved accuracy relative to several alternative emulators, particularly in lower-information settings, while maintaining near-nominal interpolative coverage. We also illustrate the method on a radiation-hydrodynamics example from Cassio, where it is used to extrapolate a derived wave-front diagnostic beyond the finest observed simulation. These results suggest that spatially regularized resolution extrapolation can provide a useful statistical tool for studying high-fidelity behavior when direct simulation is expensive.
\end{abstract}


\section{Introduction}

Radiation transport through heterogeneous media is a fundamental challenge in radiation-hydrodynamics (rad-hydro) modeling. In many relevant settings---including astrophysical transients and inertial confinement fusion---the materials encountered by radiation are not homogeneous but instead contain clumpy or fragmented substructures that cannot be directly resolved on simulation grids. These subgrid-scale heterogeneities can strongly affect radiation flow and energy deposition, yet are often represented only indirectly through heuristic or transport-based approximations, such as effective-opacity descriptions of average attenuation behavior.

Recent advances in computation and data-driven modeling present an opportunity to improve upon such approximations by learning how simulation output changes with resolution. Rather than replacing the underlying physics solver, one can use a collection of lower-resolution simulations to infer higher-resolution behavior for quantities of interest such as radiation energy, temperature, or wave-front location. This is especially appealing when the highest-fidelity simulations are computationally expensive or infeasible.

In this work, we present a Bayesian approach for \emph{spatially regularized extrapolation} of scalar fields from lower-resolution simulations. Our goal is to infer high-fidelity spatial fields using a biased and noisy set of lower-resolution observations while preserving spatial structure and quantifying predictive uncertainty. Although the examples considered here focus on two-dimensional fields, the framework extends naturally to higher-dimensional settings, including spatiotemporal problems.

Our approach is based on nonlinear growth models whose parameters are linked across space through Gaussian process (GP) priors. Using synthetic experiments designed to mimic resolution-dependent bias in radiation transport simulations, we show that the method can recover the underlying high-resolution field, provide calibrated uncertainty estimates, and support extrapolation beyond the range of observed resolutions.

\section{Background and Motivation}

A broad class of approaches to this problem is based on surrogate modeling, in which a statistical model is trained to approximate the input--output behavior of an expensive simulation code using a limited ensemble of runs \citep{yuen2019surrogate}. Gaussian process (GP) regression is especially attractive in this setting because it provides a flexible Bayesian framework for emulating simulation output with uncertainty quantification \citep{lamminpaa2025forward}. Multi-output extensions, including co-kriging and related multi-fidelity formulations, allow discrepancies between coarse- and fine-resolution runs to be modeled directly.

Related ideas also appear in multiscale and multi-fidelity methods, where coarse and fine simulations are combined through hierarchical or autoregressive constructions \citep{oughton2016hierarchical}. These methods treat lower-fidelity output as informative but biased, and then learn a correction toward the higher-fidelity response. Similar goals underlie learned super-resolution approaches \citep{zhang2025ai, exalearn2024surrogate}. Because extrapolation beyond the observed resolution range is inherently uncertain, uncertainty quantification is central to these methods, although direct GP inference can become computationally difficult for large spatial outputs, motivating scalable approximations such as the Scaled Vecchia approach of \citet{katzfuss2022scaled}.

These ideas are closely connected to the broader problem of representing subgrid effects. One line of work approaches this through asymptotic and homogenization arguments. For example, \citet{prinja2005grey} derive asymptotic solutions to the grey radiative transfer equation in binary stochastic media with material temperature coupling, showing how unresolved material variability can be represented through effective large-scale transport behavior in optically thick regimes. A complementary line of work uses ensembles of simulations in random heterogeneous media to learn how unresolved structure affects macroscopic observables. \citet{fryer2023understanding}, for instance, study radiation flow through stochastic materials and investigate surrogate relationships between summaries of subgrid structure and integrated responses such as flux. This perspective is closely aligned with the present work, where the goal is to characterize how unresolved heterogeneity induces resolution-dependent bias in resolved output.

For expensive stochastic simulations, experimental design and dimension reduction can also play important supporting roles. Nested Latin hypercube designs provide one strategy for coordinating samples across fidelity levels \citep{chen2017flexible}, while replication-based sequential design can improve emulator accuracy in regions where stochastic variability is large \citep{binois2019replication}. For high-dimensional outputs, latent-basis representations provide an additional reduction step by expressing the field in a lower-dimensional basis prior to emulation (e.g., \citet{higdon2008computer} uses a principal component basis and models the basis weights).

\section{Methods}

\subsection{Modeling Resolution-Dependent Bias}

Let \( y_{ij}(r) \) denote the observed value at pixel \((i, j)\) from a simulation run at resolution \( r \in [0,1] \), where \( r = 1 \) corresponds to the highest-fidelity simulation. We assume that each pixel’s observation across resolutions follows a nonlinear growth model:
\[
f(r; \theta_{ij}) = f_{\min, ij} + \frac{B_{ij}}{k_{ij}} \left(1 - e^{-k_{ij}(r - r_0)} \right),
\]
where \( \theta_{ij} = (f_{\min, ij}, B_{ij}, k_{ij}) \) are the pixel-specific parameters to be estimated, and \( r_0 \) is a fixed minimum resolution (here taken as zero). The model reflects the idea that the predicted value increases with resolution, eventually saturating as the resolution becomes sufficiently high.

\subsection{Bayesian Model Formulation}

To capture both parameter uncertainty and spatial correlations across pixels, we adopt a fully Bayesian formulation. Each parameter of the nonlinear growth curve
\[
\theta_{ij} = (f_{\min,ij}, B_{ij}, k_{ij})
\]
is modeled as a latent spatial field, each with its own lengthscale and process variance. Specifically, we place Gaussian process (GP) priors on the three fields
\[
f_{\min,ij} \sim \mathcal{GP}(0, K^{(f)}), \quad
B_{ij} \sim \mathcal{GP}(0, K^{(B)}), \quad
\log k_{ij} \sim \mathcal{GP}(0, K^{(k)}).
\]

Because the data are observed on a rectangular grid of pixel locations, the covariance naturally factors into Kronecker products:
\[
K^{(p)} = K^{(p)}_x \otimes K^{(p)}_y, \quad p \in \{f,B,k\},
\]
where \(K^{(p)}_x\) and \(K^{(p)}_y\) are one-dimensional covariance matrices defined along the horizontal and vertical axes of the grid. This separability is a direct consequence of the gridded design and allows efficient computation using Kronecker algebra. Each one-dimensional kernel is taken to be a squared exponential kernel, parameterized by lengthscales \(\ell_x,\ell_y\) and marginal variance \(\sigma_f^2\).

Observed simulation outputs at resolution \(r_t\) are linked to the latent fields through the nonlinear growth model:
\[
y_{ij}(r_t) \;\sim\; \mathcal{N}\!\left(f(r_t; \theta_{ij}), \; \sigma^2\right),
\]
with a shared noise variance \(\sigma^2\). This construction permits both interpolation across intermediate resolutions and extrapolation to the highest-fidelity case (\(r=1\)).

As an alternative to the Bayesian formulation, one can also fit the nonlinear growth model using penalized nonlinear least squares with spatial penalties. This provides faster approximate inference at the cost of fully accounting for uncertainty. We summarize this approach in the Appendix.

\subsection{Posterior Sampling}

Posterior inference proceeds via Markov chain Monte Carlo (MCMC), using a combination of Gibbs and Metropolis–Hastings (MH) updates:

\begin{enumerate}
  \item Update the noise variance \(\sigma^2\) using a conjugate inverse-gamma step.
  \item For each latent field \(f_{\min}, B, \log k\):
  \begin{enumerate}
    \item Propose a new field realization using a preconditioned Crank–Nicolson (pCN) scheme, which preserves the GP prior and improves mixing.
    \item Accept or reject the proposed field via an MH step based on the likelihood contribution from all resolutions.
    \item Update the GP hyperparameters (lengthscales \(\ell_x,\ell_y\), variance \(\sigma_f^2\)) with log-random walk MH updates under the priors described in the Supplement.
  \end{enumerate}
  \item For the \(\log k\) field, sample the precision parameter \(\lambda_k\) governing the soft-box penalty (Section~\ref{sec:priors}) via a Gibbs step from its gamma conditional.
\end{enumerate}

This blocked sampler maintains detailed balance and exploits the Kronecker factorization of the GP covariance for efficient computation, scaling linearly with the grid size in each dimension.

\section{Simulation Study}

To evaluate the proposed method, we design a synthetic framework in which we first add resolution-dependent biases to underlying surfaces and then inject noise. In more detail, the simulation procedure is as follows. A latent ``true'' spatial field is generated on a two-dimensional \(20 \times 20\) grid by drawing a realization from a Gaussian process (GP) with a squared exponential kernel. Independently, a second GP realization is drawn to serve as a structured bias field. This construction reflects the idea that lower-resolution simulations systematically misrepresent fine-scale heterogeneity in ways that vary smoothly across space. 

For a given resolution level \( r \in [0,1] \), the observed field is defined as
\[
Y(r) \;=\; \big(1 - \alpha(r)\big)\, \text{True Field} \;+\; \alpha(r)\, \text{Bias Field} \;+\; \varepsilon,
\]
where the bias weight is specified as \(\alpha(r) = (1-r)^{\beta}\) with exponent \(\beta \geq 1\). At the highest resolution (\(r=1\)), the bias becomes zero and the field coincides with the truth, while at lower resolutions the bias dominates. The larger $\beta$ is, the quicker the field converges to the truth. The error term \(\varepsilon\) consists of independent Gaussian noise with standard deviation \(\sigma\), representing stochastic variation or numerical error.

We vary several parameters to create a collection of test cases. The bias exponent is chosen from values \(\beta \in \{1.5, 2, 2.5\}\), so that the persistence of bias with increasing resolution ranges from low to moderate. The noise standard deviation is taken to be either \(\sigma = 0.01\), \(\sigma = 0.05\), or \(\sigma = 0.1\), allowing assessment of low, medium, and high noise scenarios. 

\begin{figure}[htbp]
    \centering
    \includegraphics[width=0.85\textwidth,trim=0 80 0 80,clip]{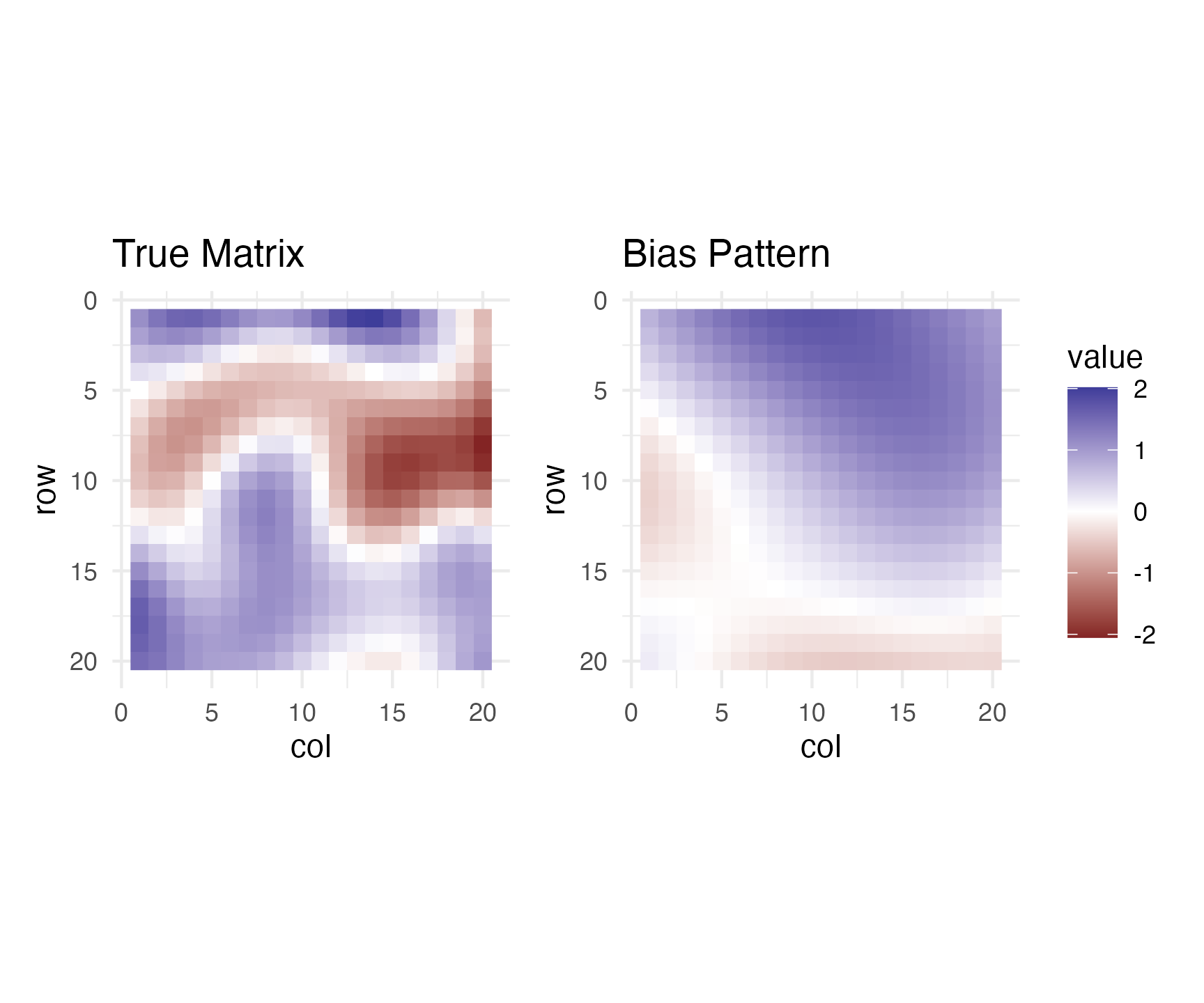}
    \caption{The true matrix (left) and the bias pattern (right) samples for a single example simulation.}
    \label{fig:sim_example}
\end{figure}

\begin{figure}[htbp]
    \centering
    \includegraphics[width=0.55\textwidth]{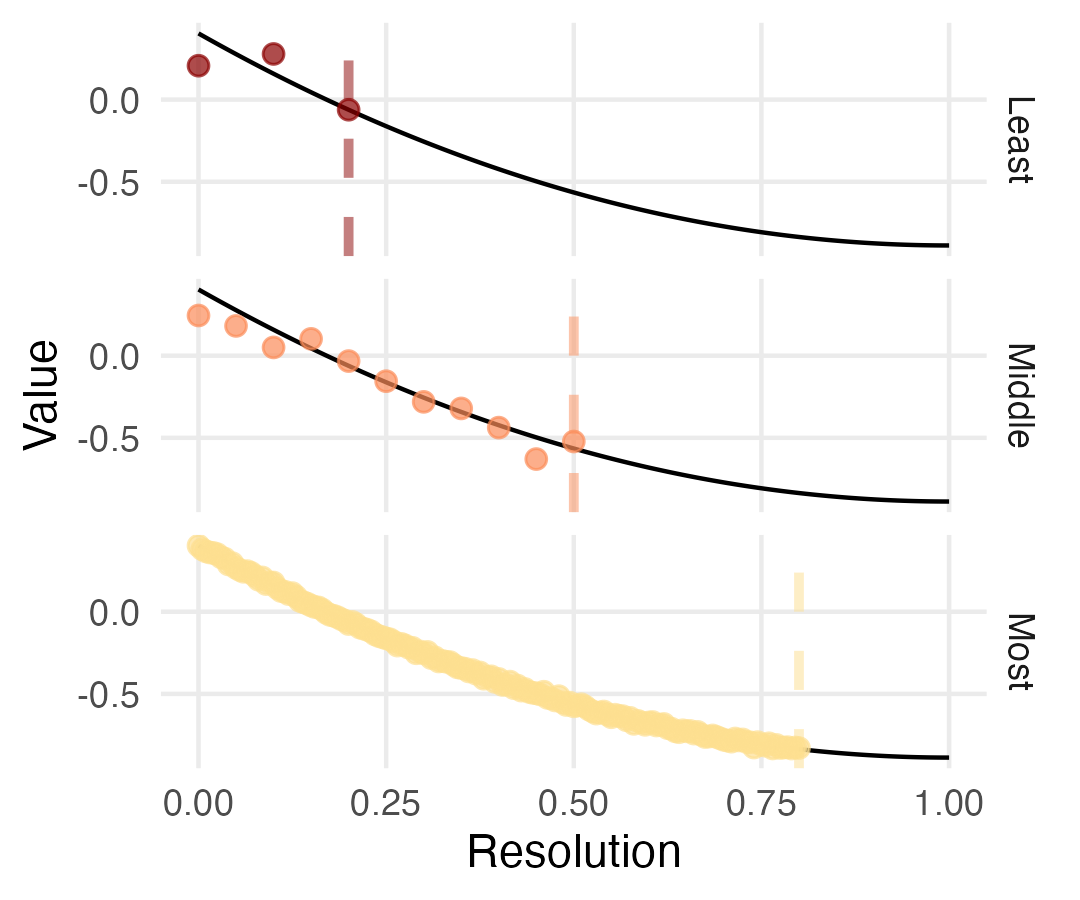}
    \caption{Example of sampled values at entry [5, 7] across resolutions for the example simulation from Figure \ref{fig:sim_example} with $\beta = 2$. As resolution increases the function resolves toward its true unbiased value. The subplots show different amounts of training information from least to most, i.e. different simulation settings. Top: max observed resolution of 0.2, low point density, $\sigma=0.1$. Middle: max observed resolution of 0.5, medium point density, $\sigma=0.05$. Bottom:  max observed resolution of 0.8, high point density, $\sigma=0.01$.}
    \label{fig:sim_example_dat}
\end{figure}

The Gaussian process governing the true field has lengthscale 0.2, while the GP for the bias field has lengthscale 0.5. Both have marginal variance 1.0. Training data consists of subsets of the resolution grid \(\{0.000, 0.005, 0.010, \ldots, 1.000\}\), with high (every element), medium (every 10th element), or low (every 20th element) point density and with the maximum observed resolution ranging between 0.2 and 0.8. This design produces synthetic datasets with realistic spatial structure, controllable noise, and varying degrees of resolution-dependent bias. An example simulation and data realizations at a single grid cell are shown in Figures \ref{fig:sim_example} and \ref{fig:sim_example_dat}.

We benchmark our Bayesian emulator against three alternative approaches. The first is a PCA+GAM method, which applies principal component analysis for dimensionality reduction followed by generalized additive models to capture resolution dependence. The second is PCA+NNET, in which the principal component scores are modeled as nonlinear functions of resolution using feedforward neural networks. The third is a fully per-cell approach, in which cells are fitted with independent generalized additive models. These competitors span a range of modeling philosophies, from flexible but non-spatial regressions to low-rank methods that impose global structure.

This simulation framework enables evaluation of both interpolative prediction, where target resolutions lie within the training range, and extrapolative prediction, where the goal is to recover the high-fidelity truth at \(r=1\).

\subsection{Recovery of the Target}

We assess performance of all methods using mean squared error (MSE) at the high-fidelity target, and also report our model's interpolative coverage of pointwise \(95\%\) credible intervals. For a prediction grid with entries \(\hat{Y}_{ij}(r)\) and target field \(Y_{ij}(r)\), MSE is computed as 
\[
\mathrm{MSE}(r) \;=\; \frac{1}{mn}\sum_{i=1}^{m}\sum_{j=1}^{n}\bigl\{\hat{Y}_{ij}(r)-Y_{ij}(r)\bigr\}^{2},
\]
with primary focus on the extrapolative case \(r=1.0\). Coverage is defined as the fraction of observed pixels whose \(95\%\) credible intervals contain the true value, with nominal coverage indicated by a horizontal reference at 0.95. 

\begin{figure}[htbp]
    \centering
    \includegraphics[width=0.95\textwidth]{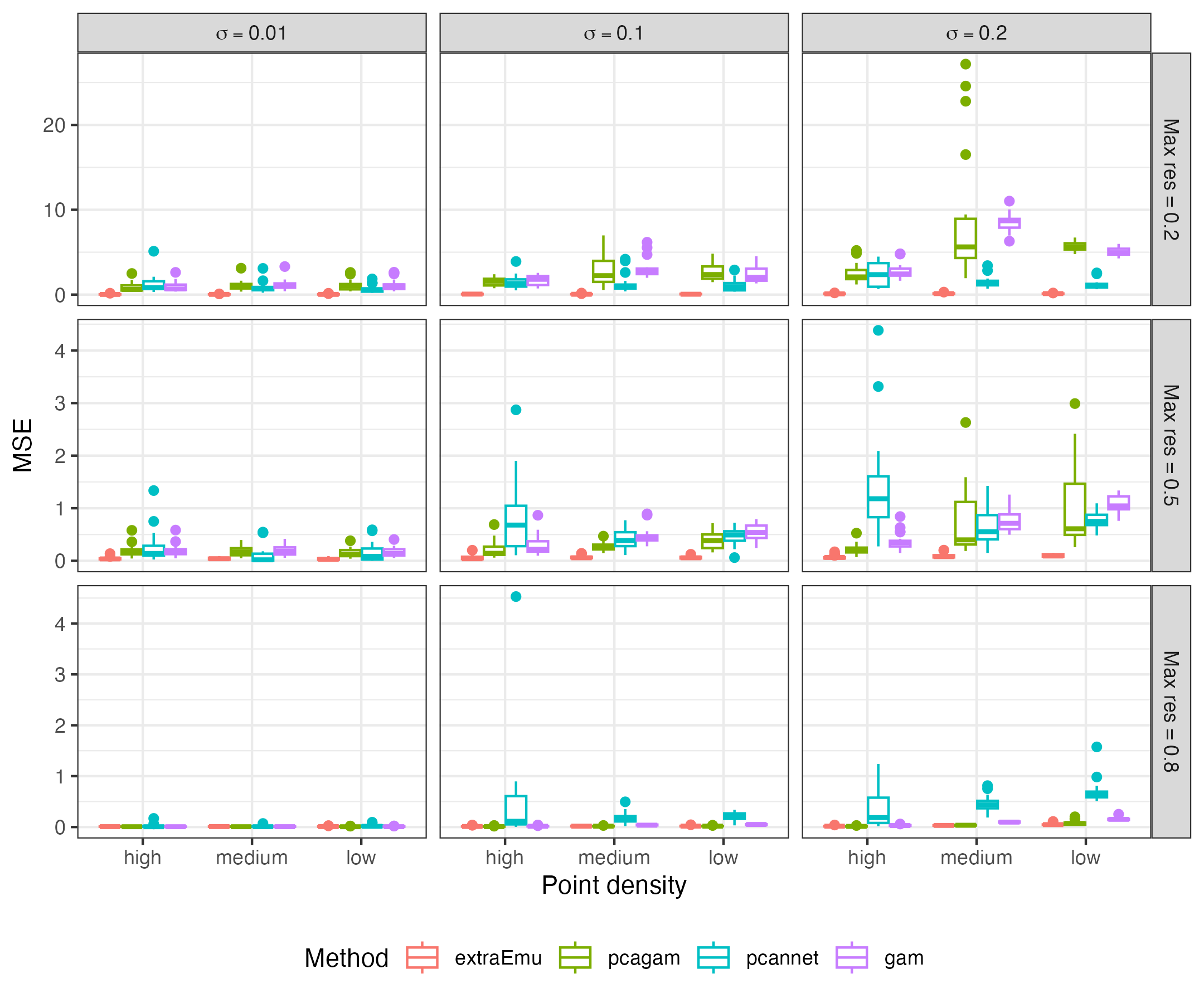}
    \caption{MSE across methods (extraEmu, PCA+GAM, PCA+NNET, per-cell GAM) for predicting at \(r=1.0\) under simulations having $\beta=2.0$. Boxplots summarize variability across replicates, stratified by maximum observed resolution, point density, and noise level \(\sigma\). Shown are the median (horizontal line in each box) and interquartile range (the box spanning the 25th to 75th percentiles), vertical lines that extend to points within 1.5 times the IQR for each method and setting, and any outliers.}
    \label{fig:mse_all}
\end{figure}

\begin{figure}[htbp]
    \centering
    \includegraphics[width=0.8\textwidth]{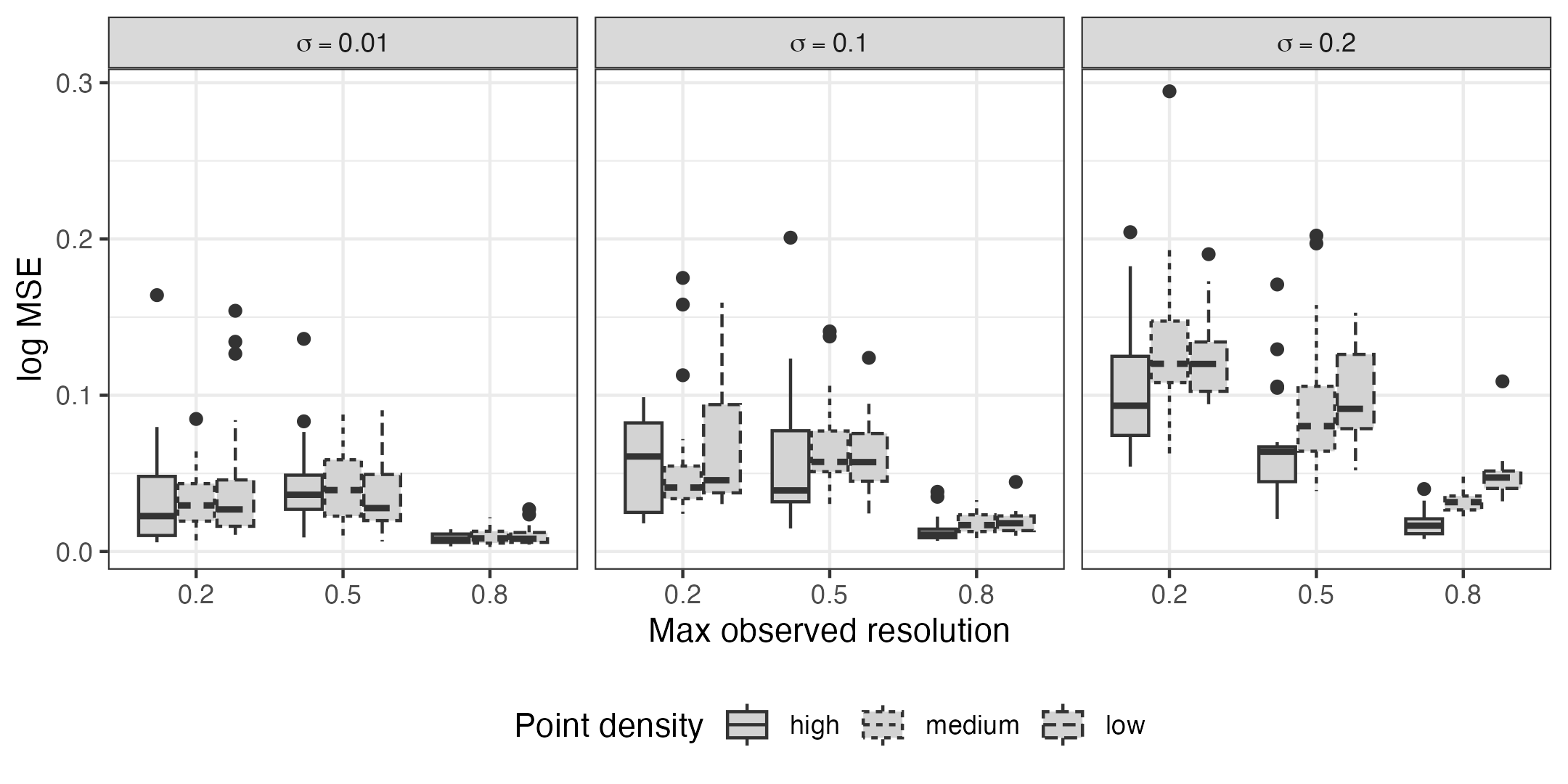}
    \caption{MSE for extraEmu for predicting at \(r=1.0\) under simulations having $\beta=2.0$. Shown are the median (horizontal line in each box) and interquartile range (the box spanning the 25th to 75th percentiles), vertical lines that extend to points within 1.5 times the IQR for each method and setting, and any outliers. As expected, MSE lowers with denser training points and with higher max observed resolution.}
    \label{fig:mse_extraemu_log}
\end{figure}

Tables \ref{tab:mse_beta1.5} through \ref{tab:mse_beta2.5} show the average MSE achieved by each method under each scenario.
Figures~\ref{fig:mse_all} and \ref{fig:mse_extraemu_log} visually summarize the predictive performance results for the 20 replicate simulations performed under each setting for $\beta=2.0$, with the former showing all methods and the latter focusing on extraEmu. Analogous figures for the bias exponent \(\beta\) values 1.5 and 2.5 are shown in the appendix.
Figure~\ref{fig:coverage_interp} shows interpolative coverage for extraEmu at \(r=0.1\), demonstrating that uncertainty intervals remain close to nominal across settings. 
Overall, extraEmu achieves consistently lower or comparable MSE relative to competitors, while maintaining near-nominal coverage at observed resolutions.
For extraEmu, MSE improves as max observed resolution and point density increase, as we would expect. 
The method is robust to different noise levels.

\begin{table}[htpb]
\centering
\caption{Average MSE by point density, maximum resolution, and noise sd for the $\beta$ = 1.5 simulations, i.e. the simulations for which the bias resolves less quickly. Smallest per group in bold.}
\centering
\begin{tabular}[t]{cccrrrr}
\toprule
Point density & Max res & $\sigma$ & extraEmu & pcagam & pcannet & gam\\
\midrule
\midrule
 &  & 0.01 & \textbf{0.03} & 0.26 & 1.22 & 0.29\\
 & 0.2 & 0.10 & \textbf{0.06} & 0.58 & 1.35 & 0.62\\
 &  & 0.20 & \textbf{0.11} & 1.68 & 2.15 & 1.84\\
\addlinespace
 &  & 0.01 & \textbf{0.01} & 0.07 & 0.57 & 0.08\\
high & 0.5 & 0.10 & \textbf{0.01} & 0.08 & 0.40 & 0.12\\
 &  & 0.20 & \textbf{0.02} & 0.12 & 1.01 & 0.17\\
\addlinespace
 &  & 0.01 & \textbf{0.01} & 0.01 & 0.04 & 0.01\\
 & 0.8 & 0.10 & \textbf{0.01} & 0.01 & 0.91 & 0.02\\
 &  & 0.20 & \textbf{0.01} & 0.01 & 0.61 & 0.03\\
\midrule
 &  & 0.01 & \textbf{0.07} & 0.25 & 0.80 & 0.28\\
 & 0.2 & 0.10 & \textbf{0.12} & 1.86 & 1.16 & 2.19\\
 &  & 0.20 & \textbf{0.16} & 5.32 & 1.52 & 7.82\\
\addlinespace
 &  & 0.01 & \textbf{0.01} & 0.07 & 0.21 & 0.09\\
medium & 0.5 & 0.10 & \textbf{0.02} & 0.24 & 0.59 & 0.30\\
 &  & 0.20 & \textbf{0.05} & 0.52 & 0.67 & 0.58\\
\addlinespace
 &  & 0.01 & \textbf{0.01} & 0.01 & 0.02 & 0.01\\
 & 0.8 & 0.10 & \textbf{0.01} & 0.01 & 0.14 & 0.03\\
 &  & 0.20 & \textbf{0.02} & 0.06 & 0.40 & 0.08\\
\midrule
 &  & 0.01 & \textbf{0.09} & 0.30 & 0.78 & 0.29\\
 & 0.2 & 0.10 & \textbf{0.14} & 1.57 & 1.28 & 1.33\\
 &  & 0.20 & \textbf{0.18} & 5.00 & 1.53 & 4.48\\
\addlinespace
 &  & 0.01 & \textbf{0.01} & 0.08 & 0.21 & 0.09\\
low & 0.5 & 0.10 & \textbf{0.02} & 0.31 & 0.71 & 0.38\\
 &  & 0.20 & \textbf{0.07} & 1.04 & 1.05 & 0.97\\
\addlinespace
 &  & 0.01 & \textbf{0.01} & 0.01 & 0.01 & 0.01\\
 & 0.8 & 0.10 & \textbf{0.01} & 0.02 & 0.20 & 0.05\\
 &  & 0.20 & \textbf{0.03} & 0.08 & 0.56 & 0.12\\
\bottomrule
\bottomrule
\end{tabular}
\label{tab:mse_beta1.5}
\end{table}

\begin{table}[htpb]
\centering
\caption{Average MSE by point density, maximum resolution, and noise sd for the $\beta$ = 2.0 simulations, i.e. the simulations for which the bias resolves with middling speed. Smallest per group in bold.}
\centering
\begin{tabular}[t]{cccrrrr}
\toprule
Point density & Max res & $\sigma$ & extraEmu & pcagam & pcannet & gam\\
\midrule
\midrule
 &  & 0.01 & \textbf{0.04} & 0.87 & 1.22 & 0.94\\
 & 0.2 & 0.10 & \textbf{0.06} & 1.52 & 1.46 & 1.72\\
 &  & 0.20 & \textbf{0.11} & 2.49 & 2.40 & 2.73\\
\addlinespace
 &  & 0.01 & \textbf{0.04} & 0.19 & 0.26 & 0.19\\
high & 0.5 & 0.10 & \textbf{0.06} & 0.22 & 0.79 & 0.30\\
 &  & 0.20 & \textbf{0.07} & 0.22 & 1.35 & 0.36\\
\addlinespace
 &  & 0.01 & \textbf{0.01} & \textbf{0.01} & 0.02 & 0.01\\
 & 0.8 & 0.10 & \textbf{0.01} & \textbf{0.01} & 0.49 & 0.02\\
 &  & 0.20 & 0.02 & \textbf{0.01} & 0.37 & 0.03\\
\midrule
 &  & 0.01 & \textbf{0.03} & 1.09 & 0.84 & 1.19\\
 & 0.2 & 0.10 & \textbf{0.06} & 2.77 & 1.29 & 3.06\\
 &  & 0.20 & \textbf{0.13} & 8.99 & 1.50 & 8.47\\
\addlinespace
 &  & 0.01 & \textbf{0.04} & 0.18 & 0.12 & 0.20\\
medium & 0.5 & 0.10 & \textbf{0.07} & 0.27 & 0.42 & 0.46\\
 &  & 0.20 & \textbf{0.09} & 0.76 & 0.67 & 0.78\\
\addlinespace
 &  & 0.01 & \textbf{0.01} & \textbf{0.01} & 0.01 & 0.01\\
 & 0.8 & 0.10 & \textbf{0.02} & \textbf{0.02} & 0.17 & 0.04\\
 &  & 0.20 & \textbf{0.03} & 0.04 & 0.46 & 0.10\\
\midrule
 &  & 0.01 & \textbf{0.04} & 1.13 & 0.66 & 1.16\\
 & 0.2 & 0.10 & \textbf{0.07} & 2.63 & 1.06 & 2.35\\
 &  & 0.20 & \textbf{0.12} & 5.71 & 1.16 & 5.06\\
\addlinespace
 &  & 0.01 & \textbf{0.03} & 0.15 & 0.16 & 0.16\\
low & 0.5 & 0.10 & \textbf{0.06} & 0.38 & 0.47 & 0.54\\
 &  & 0.20 & \textbf{0.10} & 1.06 & 0.77 & 1.07\\
\addlinespace
 &  & 0.01 & \textbf{0.01} & \textbf{0.01} & 0.02 & 0.01\\
 & 0.8 & 0.10 & \textbf{0.02} & \textbf{0.02} & 0.20 & 0.05\\
 &  & 0.20 & \textbf{0.05} & 0.08 & 0.70 & 0.15\\
\bottomrule
\bottomrule
\end{tabular}
\label{tab:mse_beta2.0}
\end{table}

\begin{table}[htpb]
\centering
\caption{Average MSE by point density, maximum resolution, and noise sd for the $\beta$ = 2.5 simulations, i.e. the simulations for which the bias resolves more quickly. Smallest per group in bold.}
\centering
\begin{tabular}[t]{cccrrrr}
\toprule
Point density & Max res & $\sigma$ & extraEmu & pcagam & pcannet & gam\\
\midrule
\midrule
 &  & 0.01 & \textbf{0.12} & 1.76 & 0.95 & 1.87\\
 & 0.2 & 0.10 & \textbf{0.17} & 1.90 & 1.25 & 2.34\\
 &  & 0.20 & \textbf{0.21} & 2.69 & 1.52 & 3.11\\
\addlinespace
 &  & 0.01 & \textbf{0.04} & 0.20 & 0.17 & 0.21\\
high & 0.5 & 0.10 & \textbf{0.05} & 0.21 & 0.46 & 0.30\\
 &  & 0.20 & \textbf{0.08} & 0.25 & 1.35 & 0.43\\
\addlinespace
 &  & 0.01 & \textbf{0.01} & \textbf{0.01} & 0.04 & 0.01\\
 & 0.8 & 0.10 & \textbf{0.01} & \textbf{0.01} & 0.38 & 0.01\\
 &  & 0.20 & \textbf{0.01} & \textbf{0.01} & 1.00 & 0.02\\
\midrule
 &  & 0.01 & \textbf{0.13} & 1.64 & 0.75 & 1.74\\
 & 0.2 & 0.10 & \textbf{0.18} & 4.03 & 0.99 & 3.87\\
 &  & 0.20 & \textbf{0.26} & 9.90 & 1.30 & 9.53\\
\addlinespace
 &  & 0.01 & \textbf{0.03} & 0.16 & 0.19 & 0.17\\
medium & 0.5 & 0.10 & \textbf{0.09} & 0.26 & 0.41 & 0.54\\
 &  & 0.20 & \textbf{0.13} & 0.55 & 0.42 & 0.84\\
\addlinespace
 &  & 0.01 & 0.01 & \textbf{0.00} & 0.02 & 0.00\\
 & 0.8 & 0.10 & 0.02 & \textbf{0.01} & 0.16 & 0.03\\
 &  & 0.20 & \textbf{0.03} & 0.04 & 0.46 & 0.09\\
\midrule
 &  & 0.01 & \textbf{0.15} & 2.36 & 0.67 & 2.35\\
 & 0.2 & 0.10 & \textbf{0.16} & 3.44 & 0.75 & 3.14\\
 &  & 0.20 & \textbf{0.27} & 7.05 & 1.04 & 6.09\\
\addlinespace
 &  & 0.01 & \textbf{0.04} & 0.18 & 0.16 & 0.20\\
low & 0.5 & 0.10 & \textbf{0.09} & 0.40 & 0.45 & 0.56\\
 &  & 0.20 & \textbf{0.18} & 0.76 & 0.82 & 1.30\\
\addlinespace
 &  & 0.01 & 0.01 & \textbf{0.00} & 0.04 & 0.00\\
 & 0.8 & 0.10 & \textbf{0.02} & \textbf{0.02} & 0.25 & 0.05\\
 &  & 0.20 & \textbf{0.04} & 0.08 & 0.60 & 0.13\\
\bottomrule
\bottomrule
\end{tabular}
\label{tab:mse_beta2.5}
\end{table}

\begin{figure}[htbp]
    \centering
    \includegraphics[width=0.8\textwidth]{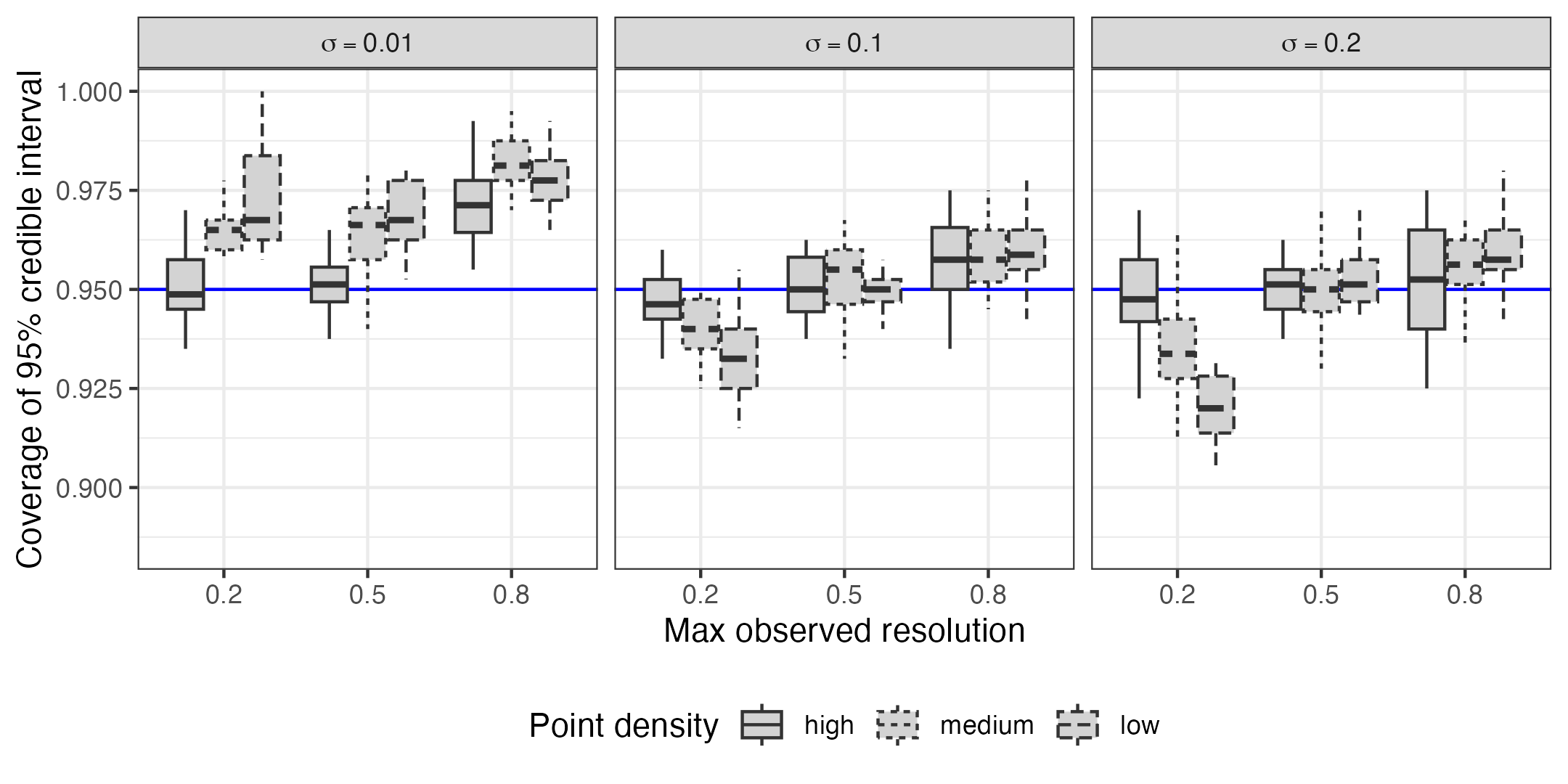}
    \caption{Data coverage of extraEmu for predicting at \(r=0.1\) under simulations having $\beta=2.0$. The subplots display the median, the interquartile range (the box spanning the 25th to 75th percentiles), and vertical lines that extend to points within 1.5 times the IQR for each and setting. Coverage is generally at or above nominal, with slightly below nominal coverage occurring in higher noise scenarios when the max observed resolution is 0.2.}
    \label{fig:coverage_interp}
\end{figure}

\begin{figure}[htbp]
    \centering
    \includegraphics[width=0.95\textwidth,trim=0 60 0 60,clip]{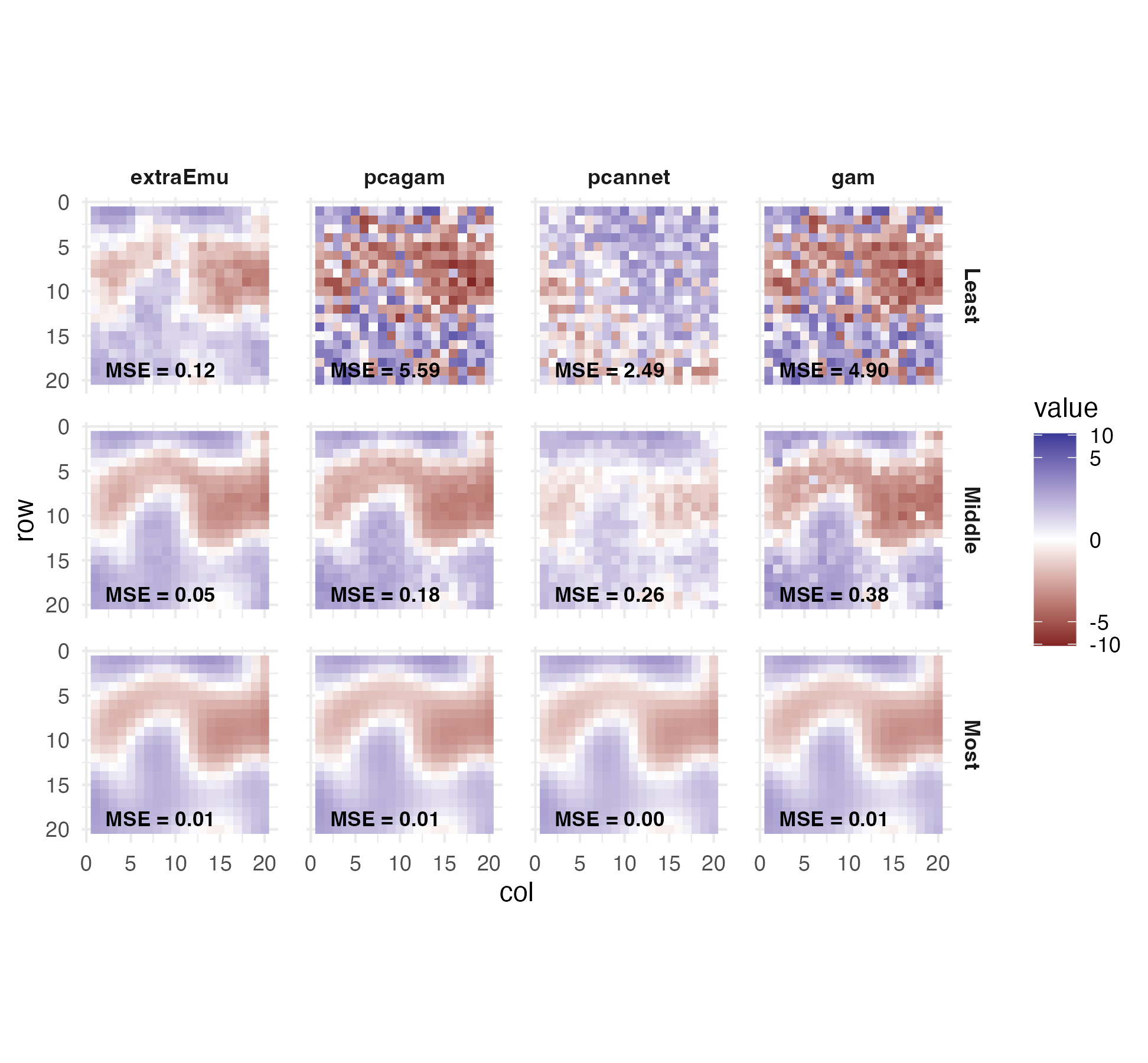}
    \caption{Results for the example simulation introduced in Figures \ref{fig:sim_example} and \ref{fig:sim_example_dat}. The extraEmu has superior performance in all but the highest information scenario, in which it has comparable performance to competitors, with the greatest relative improvement seen in the lowest information scenario.}
    \label{fig:sim_example_res}
\end{figure}

Figure \ref{fig:sim_example_res} shows the predicted high resolution matrix from our example simulation under the three illustrative scenarios. Visual examinations of the predicted vs actual fields for extraEmu confirm high fidelity recovery under each scenario, with competitor methods only able to achieve similar or slightly improved predictive error in the extreme most-information case, but suffering much worse performance in the other two lower-information scenarios.

\section{Cassio Illustration}

To illustrate the method on data from a radiation-hydrodynamics application, we consider a clump test problem calculated with Cassio. Cassio is an ICF-oriented radiation-hydrodynamics code built on the Eulerian adaptive-mesh RAGE framework and includes higher-order radiation-transport options through implicit Monte Carlo and discrete-ordinates solvers \citep{Gittings2008,Haines2022}. The setup consists of a Cartesian domain with embedded high-density clumps and a time-dependent radiation drive, producing a propagating radiation wave through a heterogeneous medium. We analyze five runs of this same problem generated at increasing spatial resolution.

\begin{figure}[hptb]
  \centering
  \includegraphics[width=0.8\textwidth]{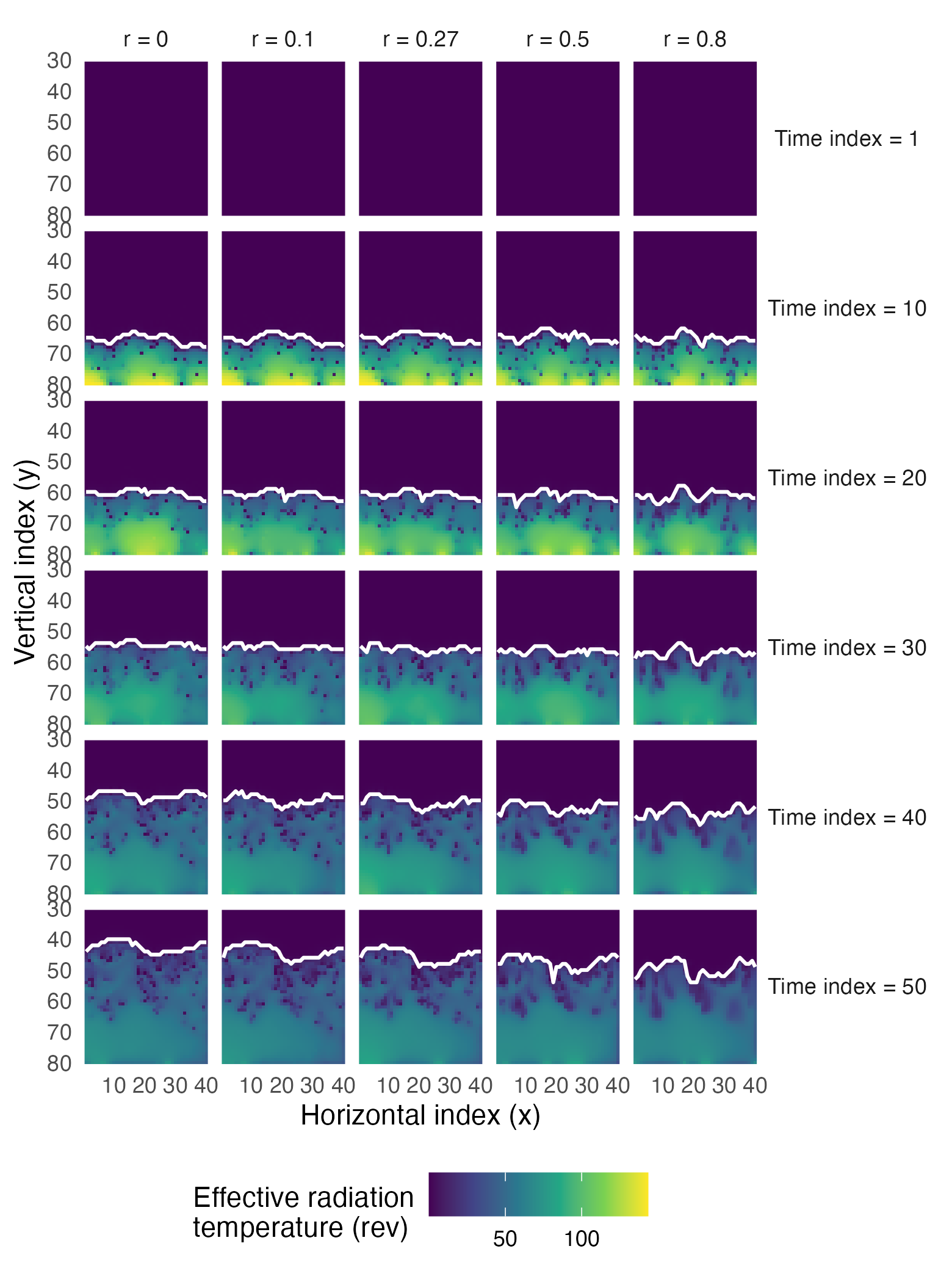}
  \caption{Effective radiation temperature (\texttt{rev}) fields at selected saved times (rows) and observed resolutions (columns). The white curve marks the radiation front location used to construct the emulator input, defined at each horizontal location as the first vertical index at which \texttt{rev} exceeds 3.}
  \label{fig:rev-front-dumps}
\end{figure}

The simulation output \texttt{rev} is the effective radiation temperature field. For each run, horizontal location, and saved time, we define the radiation front as the first vertical index at which \texttt{rev} exceeds the fixed threshold of 3 in the stored output units. Figure~\ref{fig:rev-front-dumps} shows the underlying \texttt{rev} fields and the resulting threshold-defined fronts. The overlaid curves show how the front propagates over time and how its shape changes with resolution.

Each overlaid curve forms one horizontal slice of the space--time front-location image analyzed below. Stacking these curves across all saved times produces a two-dimensional image for each resolution, with axes corresponding to horizontal position and time and pixel values corresponding to the inferred front location.
This construction fits naturally within the framework developed above. Each pixel in the resulting space--time image is treated as a scalar quantity observed across resolution levels, and the collection of pixels is linked through spatial priors so that nearby locations in space and time borrow strength from one another. For this example, the five observed runs are placed on a quadratically scaled resolution axis, and inference is used to extrapolate beyond the finest observed case to an unobserved limiting resolution. In contrast to the synthetic study, where truth is known by construction, the goal here is not formal benchmarking but rather to demonstrate that the emulator can be applied directly to a physically meaningful diagnostic extracted from a production rad-hydro workflow.

\begin{figure}[htbp]
    \centering
    \includegraphics[width=0.95\textwidth]{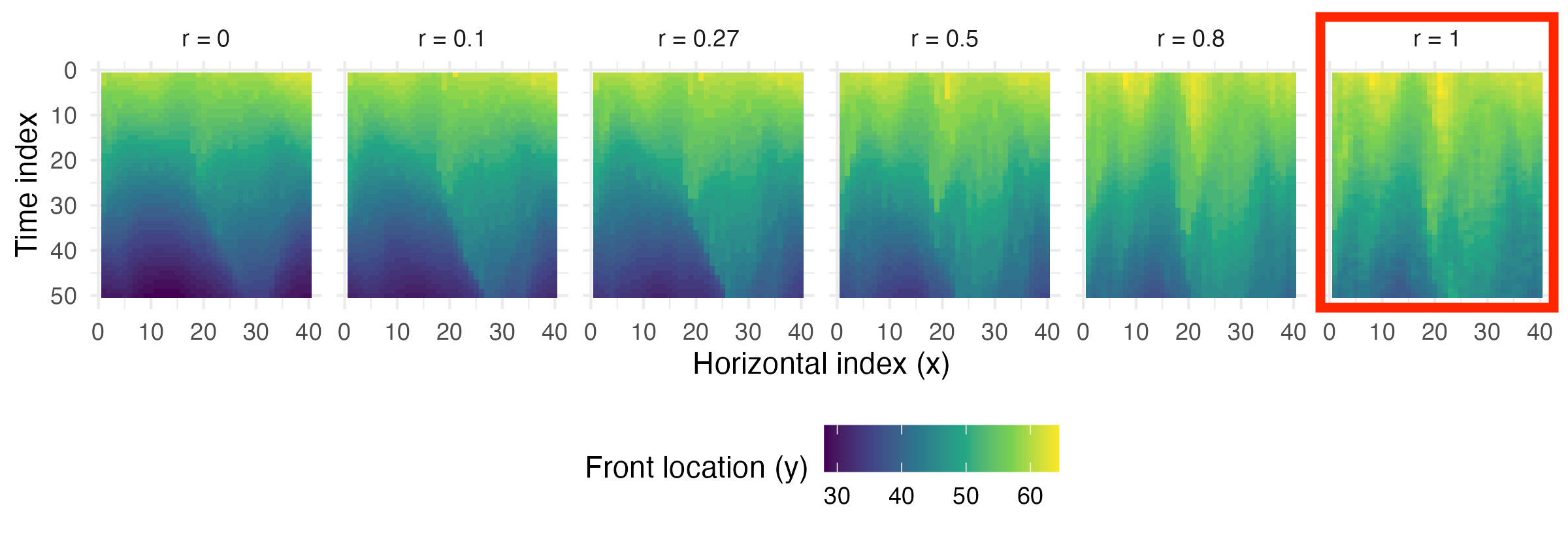}
    \caption{Wave-front location images derived from the Cassio clump-test example across the five observed resolutions, together with the extraEmu posterior mean prediction at the unobserved limiting resolution. Each panel shows the front location as a function of horizontal position and time. The final panel is highlighted to emphasize that it is obtained by extrapolation beyond the highest observed resolution.}
    \label{fig:rage_front_example}
\end{figure}

Figure~\ref{fig:rage_front_example} shows the observed wave-front images across the five available resolutions together with the extrapolated image at the limiting resolution. As resolution increases, the front location evolves in a systematic way across the space--time domain, with a petal-shaped wave pattern refining into more detailed wave fronts. The extrapolated field summarizes the model-based estimate of the limiting wave-front location while preserving the large-scale coherence visible in the observed sequence. In this setting, the extrapolated image should be interpreted as a posterior estimate of the unresolved limiting behavior, rather than as a directly validated ground truth. This example nevertheless illustrates the intended use case of the method: leveraging progressively refined simulations to infer the higher-resolution structure of a spatially gridded quantity that would otherwise require a substantially more expensive run.

\section{Discussion}

We have presented a Bayesian framework for resolution-extrapolative emulation of spatially gridded data. The method is designed for settings in which lower-resolution simulations are available, higher-resolution simulations are costly, and the primary goal is to infer the limiting field while preserving spatial structure and quantifying uncertainty. In the synthetic study, the method performs well across a range of bias, noise, and training-density settings, and in the Cassio illustration it produces a reasonable extrapolated wave-front estimate from a small sequence of progressively refined runs. Taken together, these examples support the use of the framework as a statistical complement to direct simulation, rather than as a replacement for it.

Several extensions are natural. One is to allow the resolution-response relationship itself to depend on additional simulation settings, such as material parameters, drive conditions, or geometry descriptors. In that case, the extrapolation would no longer be indexed by resolution alone, but by resolution jointly with other inputs, allowing information to be shared across families of related runs. Another important direction is extension to three-dimensional output. The present construction is written for two-dimensional fields, but the same idea could be carried to 3D by placing structured priors on parameter fields over three spatial axes, or over space and time for fully spatiotemporal problems, with additional approximation needed for computational scalability. 

The current formulation also assumes a shared residual variance across spatial locations and resolution levels. Although this provides a parsimonious starting point, it may be restrictive when numerical variability or model discrepancy changes with resolution or varies across the spatial domain. A natural extension would allow the residual variance to depend on resolution, spatial location, or both, for example by placing a structured prior on the log-variance field.

A more substantive modeling question concerns cases in which the behavior across resolution is not well described by a single smooth approach to a limiting bias field. In some applications, one may see structured intermediate-resolution effects before the output settles into a shape that is amenable to the present growth model. In such cases, a single monotone curve in resolution may be too restrictive. One possible response is to introduce a richer resolution model, for example by allowing multiple regimes, a change point, or a decomposition into a transient component plus an asymptotic component. Practically, this suggests using the current model when the observed resolution path is approximately smooth and saturating, and considering more flexible alternatives when held-out resolutions show systematic departures from that pattern. Developing such models, while retaining interpretable extrapolation and stable uncertainty quantification, is an important direction for future work.

\section*{Acknowledgments}

Research presented in this manuscript was supported by the Laboratory Directed Research and Development (LDRD) program of Los Alamos National Laboratory (LANL) under project No. 20251075ER. Work at LANL is conducted under the auspices of the United States Department of Energy. Approved for public release: LA-UR-26-22907.

\bibliographystyle{apalike}
\bibliography{background}

\newpage

\appendix
\section{Appendix}

\subsection{Bayesian Prior Specification and Justification}
\label{sec:priors}

Here we detail the priors and their motivations:

\paragraph{Gaussian Process Priors.}
Each latent parameter field receives an independent GP prior with zero mean and Kronecker-structured squared-exponential covariance. The priors on the hyperparameters are chosen as follows:
\begin{itemize}
  \item \textbf{Lengthscales} \(\ell_x,\ell_y\): the inverse lengthscales \(1/\ell_x\) and \(1/\ell_y\) receive half-\(t\) priors with three degrees of freedom and scale 3. These priors favor spatial smoothness while retaining sufficiently heavy tails to allow localized variation.
  \item \textbf{Marginal standard deviation} \(\sigma_f\): a half-normal prior with scale 5 favors shrinkage toward low-amplitude spatial variation while allowing larger excursions when supported by the data. Each one-dimensional covariance factor is scaled by \(\sigma_f\), so their Kronecker product has marginal variance \(\sigma_f^2\).
\end{itemize}

\paragraph{Soft-Box Prior on the Asymptotic Increment.}
Under the nonlinear growth model, the change between the initial value and limiting asymptote is
\[
A_{ij} = \frac{B_{ij}}{k_{ij}}
       = B_{ij}\exp(-\log k_{ij}).
\]
A quadratic soft-box penalty leaves values within \(A_{ij}\in[-2,2]\) unpenalized and increasingly penalizes values outside this interval. The penalty strength is assigned the shape--rate prior
\[
\lambda_A \sim \operatorname{Gamma}(1,0.01),
\]
allowing the data to adaptively control the degree of enforcement.

\paragraph{Noise Variance.}
Writing the observation precision as \(\tau=\sigma^{-2}\), we assign
\[
\tau \sim \operatorname{Gamma}(0.1,10^{-4}),
\]
using the shape--rate parameterization. Equivalently,
\[
\sigma^2 \sim \operatorname{Inverse\text{-}Gamma}(0.1,10^{-4}).
\]
This proper, weakly informative prior ensures positivity while allowing a broad range of residual variation.


\subsection{Additional Simulation Results}

\begin{figure}[htbp]
    \centering
    \includegraphics[width=0.85\textwidth]{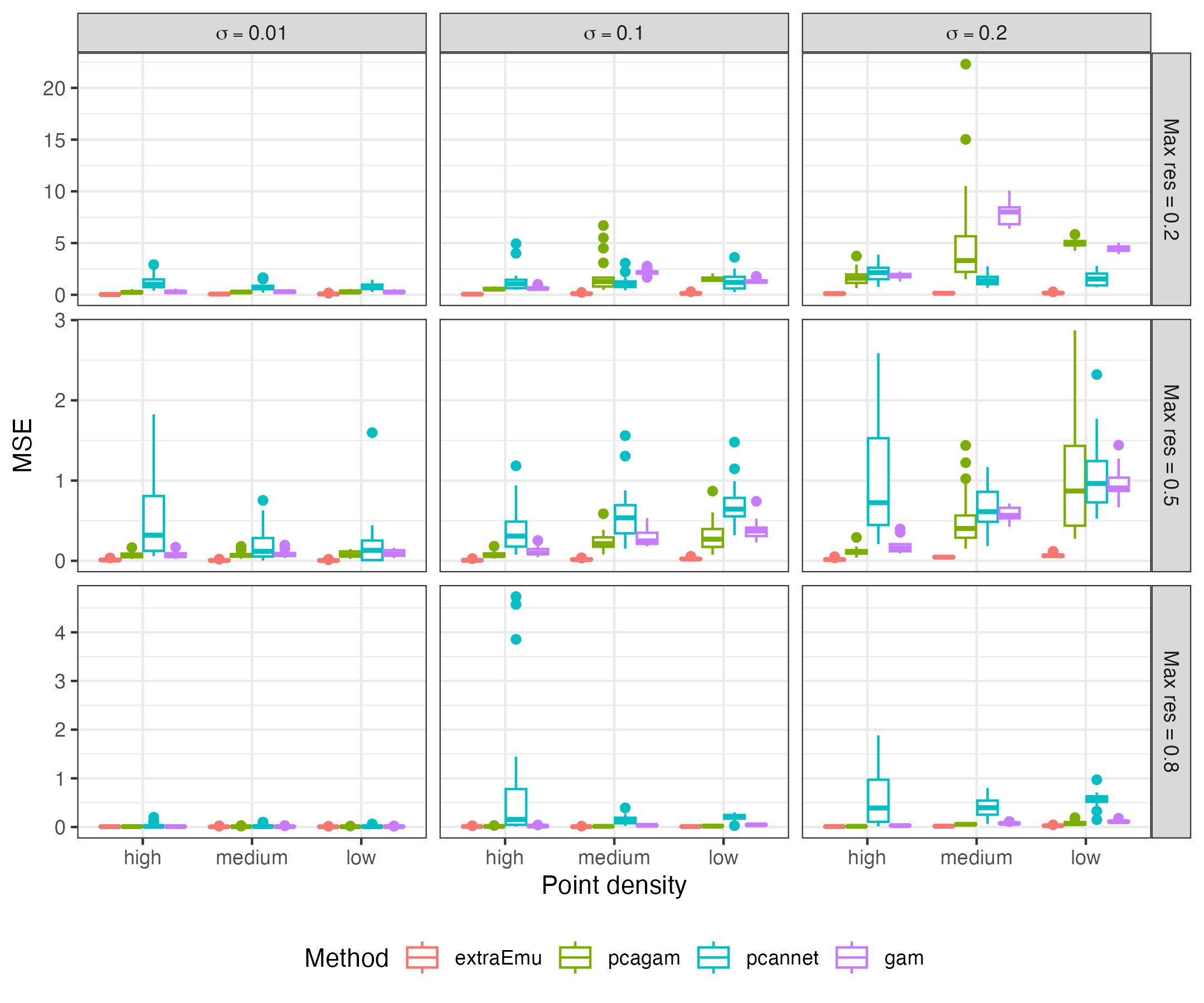}
    \caption{MSE across methods (extraEmu, PCA+GAM, PCA+NNET, per-cell GAM) for predicting at \(r=1.0\) under simulations having $\beta=1.5$. The boxplots summarize variability across replicates, stratified by maximum observed resolution, point density, and noise level \(\sigma\).}
    \label{fig:mse_all_1.5}
\end{figure}

\begin{figure}[htbp]
    \centering
    \includegraphics[width=0.8\textwidth]{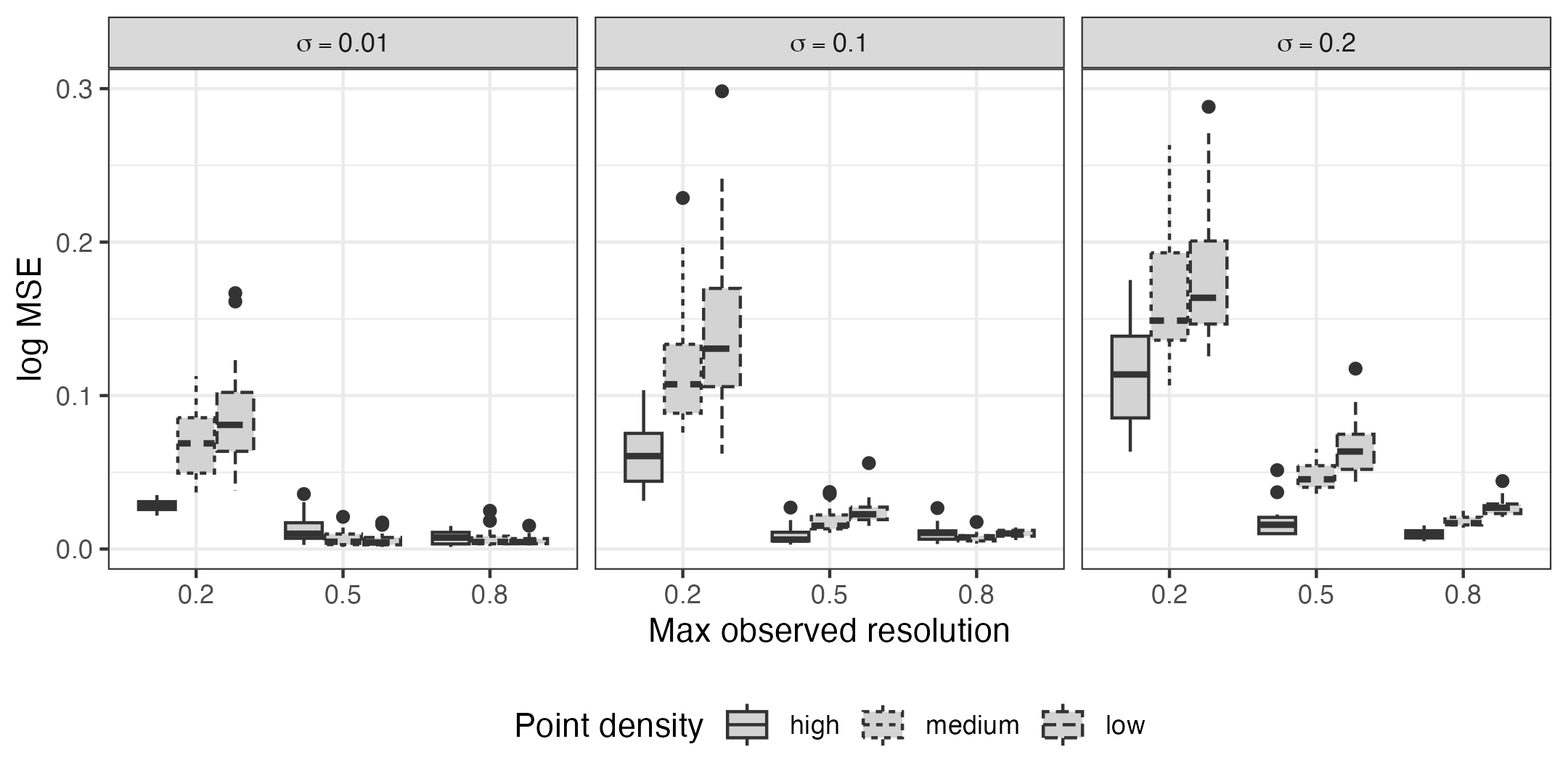}
    \caption{MSE for extraEmu for predicting at \(r=1.0\) under simulations having $\beta=1.5$. As expected, MSE lowers with denser training points and with higher max observed resolution.}
    \label{fig:mse_extraemu_log_1.5}
\end{figure}

\begin{figure}[htbp]
    \centering
    \includegraphics[width=0.85\textwidth]{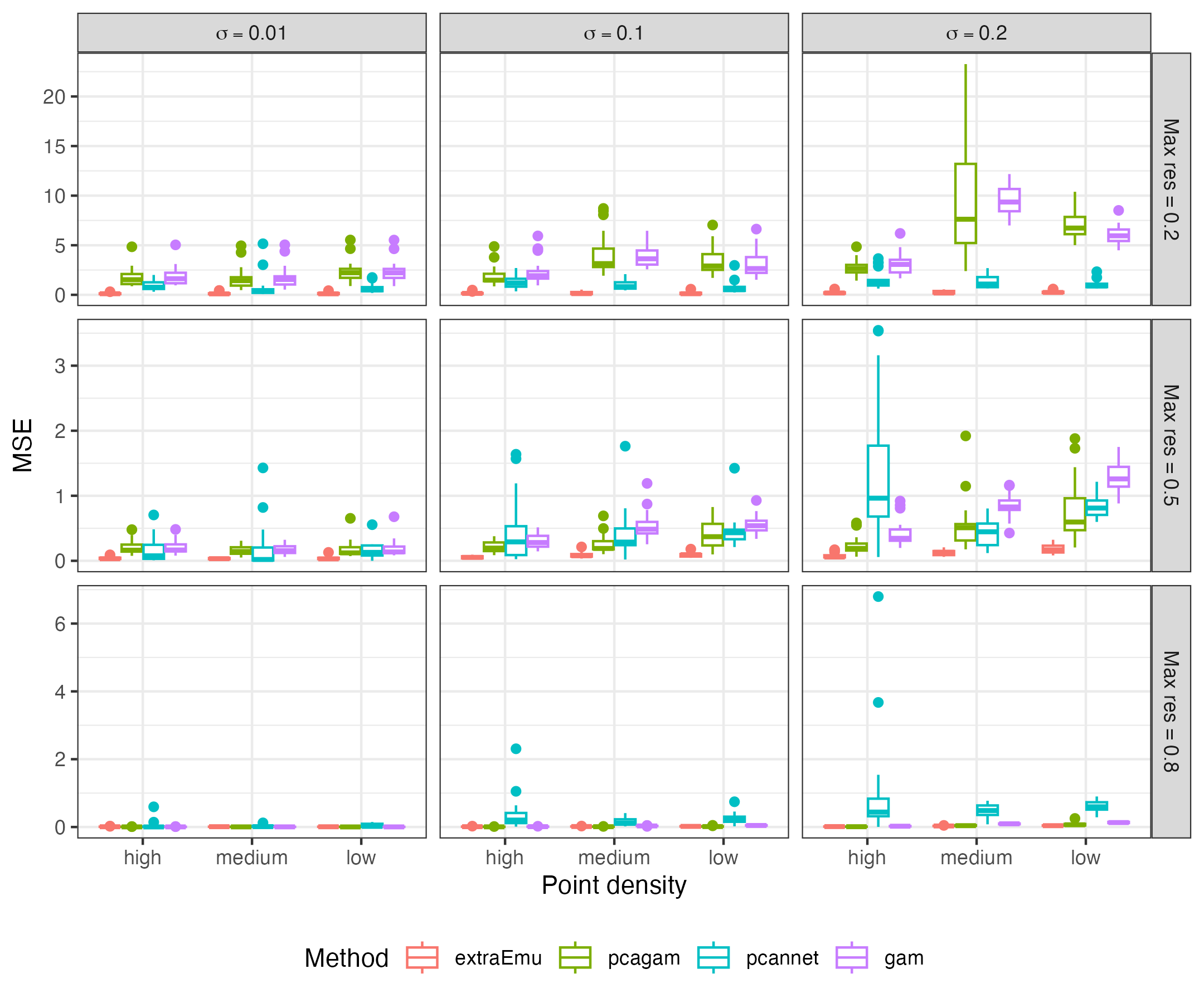}
    \caption{MSE across methods (extraEmu, PCA+GAM, PCA+NNET, per-cell GAM) for predicting at \(r=1.0\) under simulations having $\beta=2.5$. Boxplots summarize variability across replicates, stratified by maximum observed resolution, point density, and noise level \(\sigma\).}
    \label{fig:mse_all_2.5}
\end{figure}

\begin{figure}[htbp]
    \centering
    \includegraphics[width=0.8\textwidth]{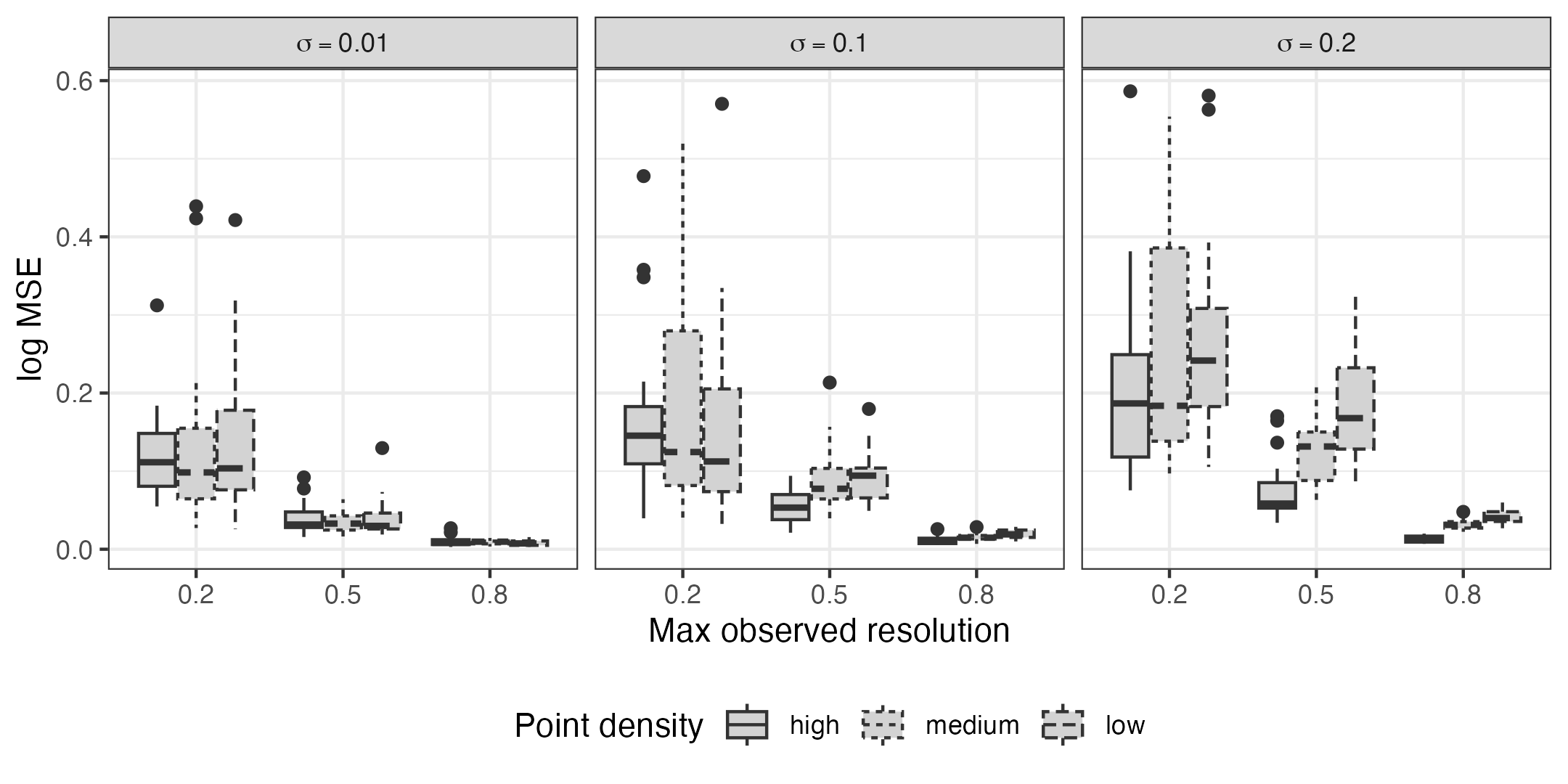}
    \caption{MSE for extraEmu for predicting at \(r=1.0\) under simulations having $\beta=2.5$. As expected, MSE lowers with denser training points and with higher max observed resolution.}
    \label{fig:mse_extraemu_log_2.5}
\end{figure}

\begin{figure}[htbp]
    \centering
    \includegraphics[width=0.8\textwidth]{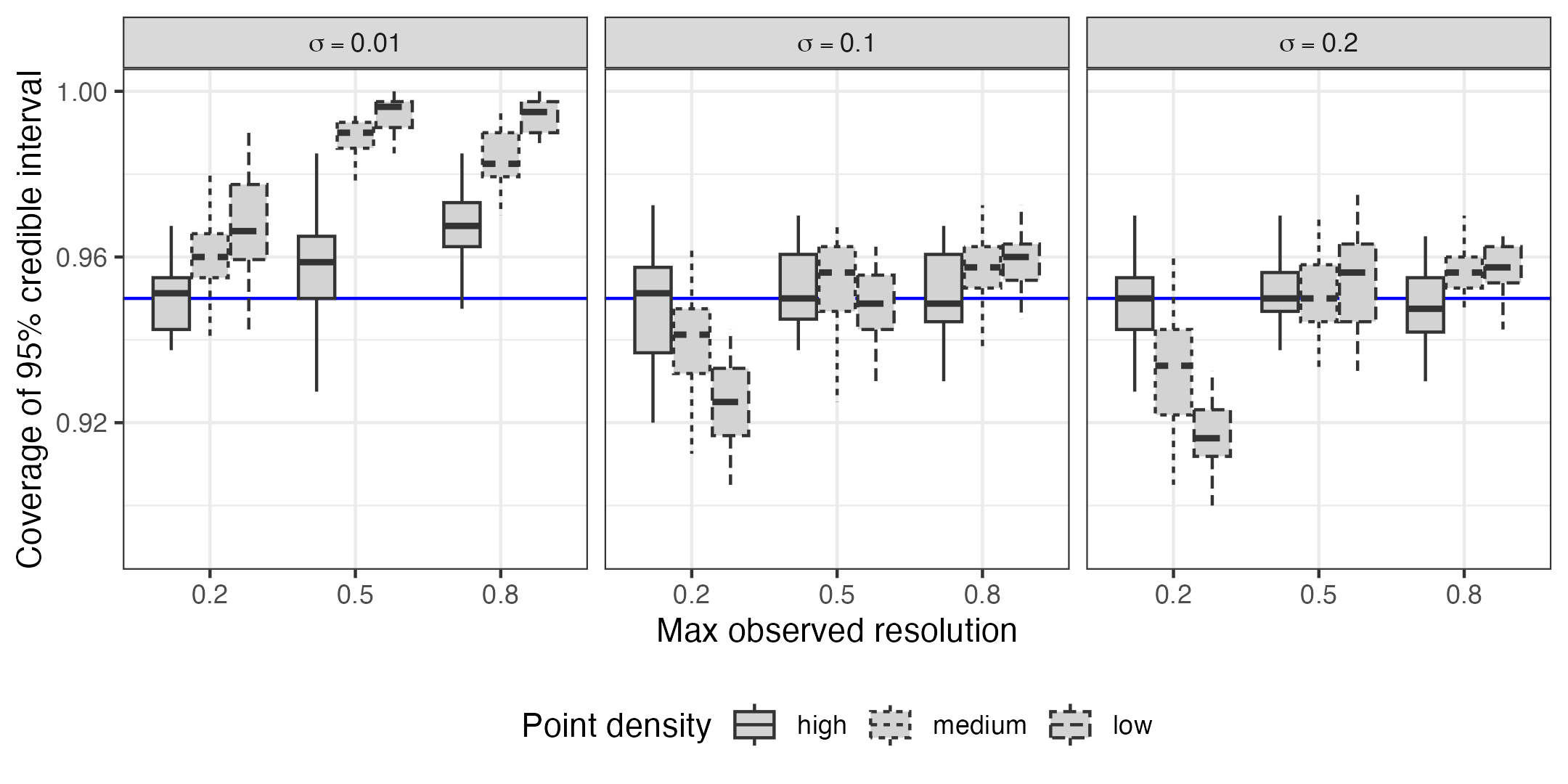}
    \caption{Data coverage of extraEmu for predicting at \(r=0.1\) under simulations having $\beta=1.5$. Coverage is generally at or above nominal, with slightly below nominal coverage occurring in higher noise scenarios when the max observed resolution is 0.2.}
    \label{fig:coverage_interp_1.5}
\end{figure}

\begin{figure}[htbp]
    \centering
    \includegraphics[width=0.8\textwidth]{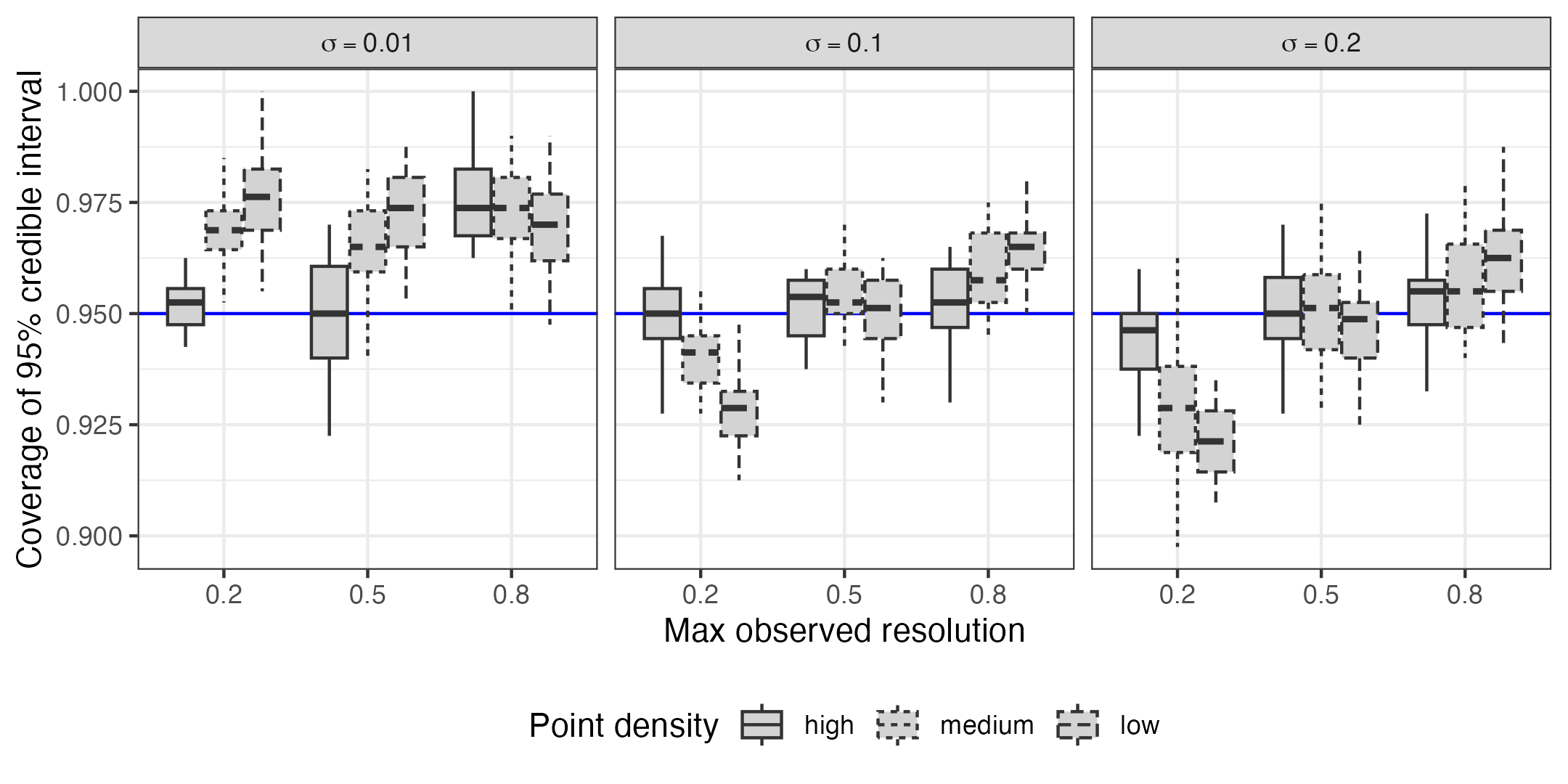}
    \caption{Data coverage of extraEmu for predicting at \(r=0.1\) under simulations having $\beta=2.5$. Coverage is generally at or above nominal, with slightly below nominal coverage occurring in higher noise scenarios when the max observed resolution is 0.2.}
    \label{fig:coverage_interp_2.5}
\end{figure}

Figures~\ref{fig:mse_all_1.5} and \ref{fig:mse_extraemu_log_1.5} visually summarize the predictive performance results and for the 20 replicate simulations performed under each setting for $\beta=1.5$; Figures~\ref{fig:mse_all_2.5} and \ref{fig:mse_extraemu_log_2.5} do so for $\beta=2.5$.
Figures~\ref{fig:coverage_interp_1.5} and \ref{fig:coverage_interp_2.5} show interpolative coverage for extraEmu at \(r=0.1\) for the 20 replicate simulations performed under each setting for $\beta=1.5$ and $\beta=2.5$, respectively.
Within each plot is shown the median (horizontal line in each box) and interquartile range (the box spanning the 25th to 75th percentiles), vertical lines that extend to points within 1.5 times the IQR for each method and setting, and any outliers.

\begin{figure}[htbp]
    \centering
    \includegraphics[width=0.95\textwidth]{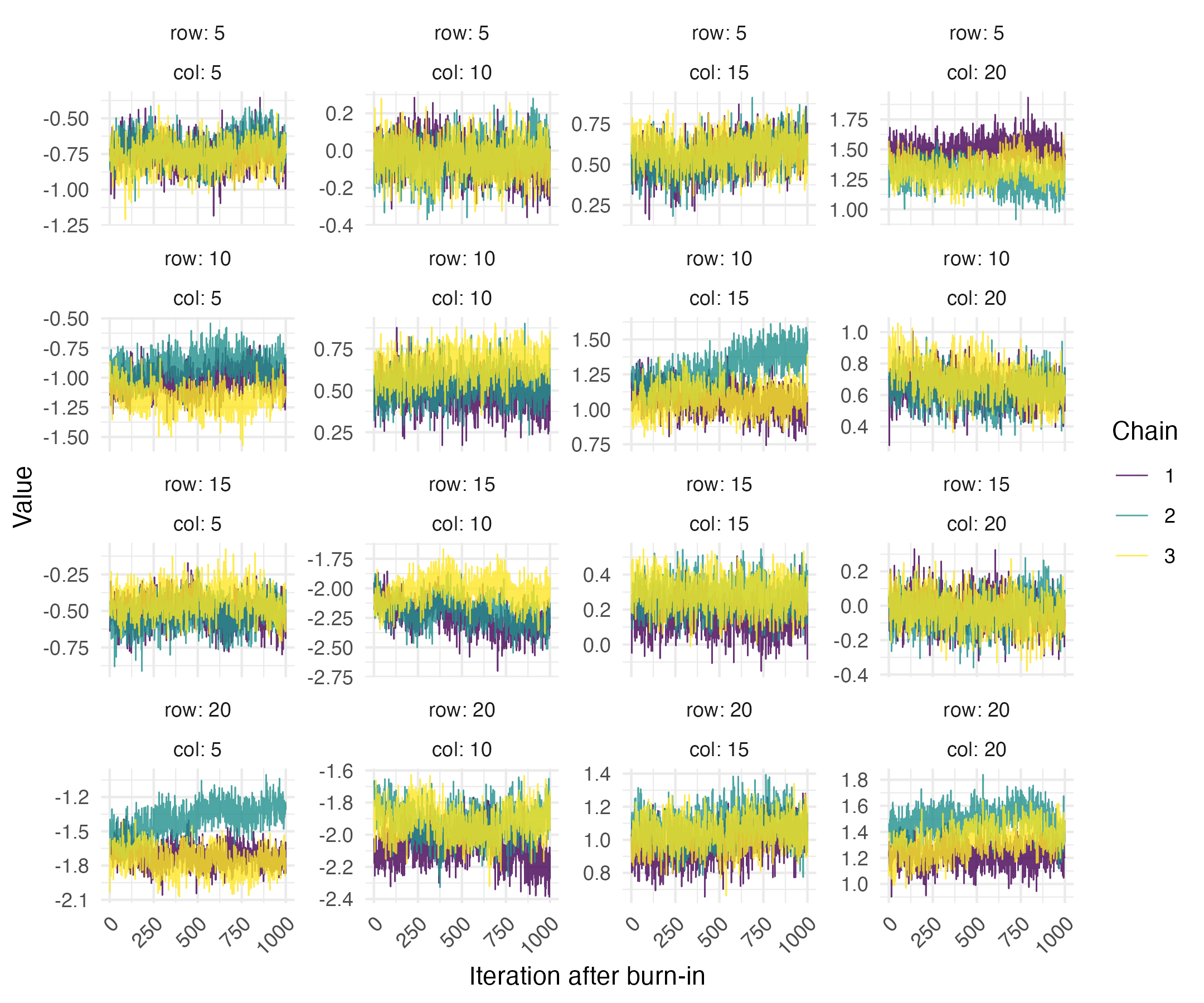}
    \caption{Trace plots of $Y^{\text{pred}}_{ij}(1)$ at select grid locations.}
    \label{fig:trace_0}
\end{figure}

\begin{figure}[htbp]
    \centering
    \includegraphics[width=0.95\textwidth]{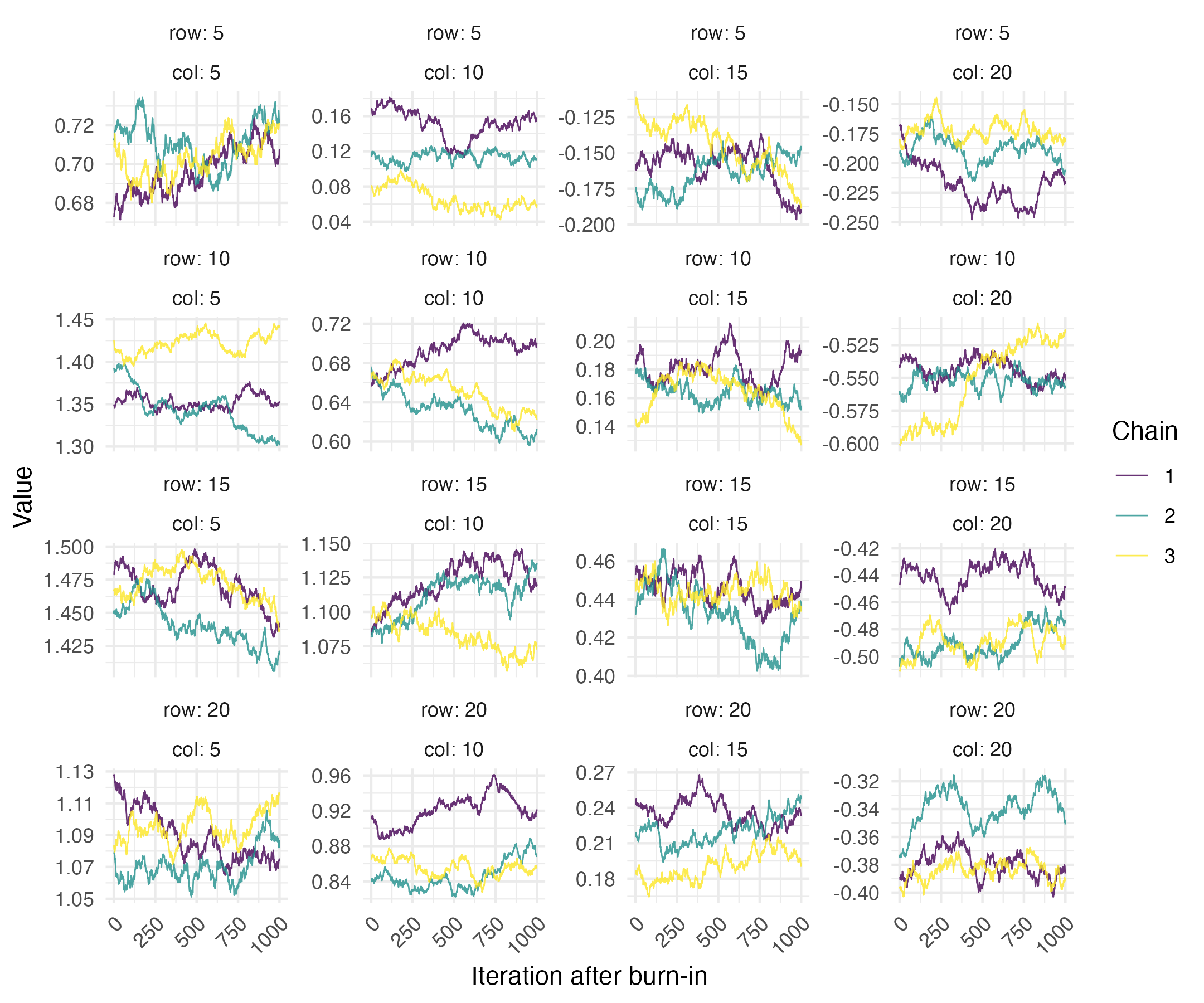}
    \caption{Trace plots of $f$ at select grid locations.}
    \label{fig:trace_1a}
\end{figure}

\begin{figure}[htbp]
    \centering
    \includegraphics[width=0.95\textwidth]{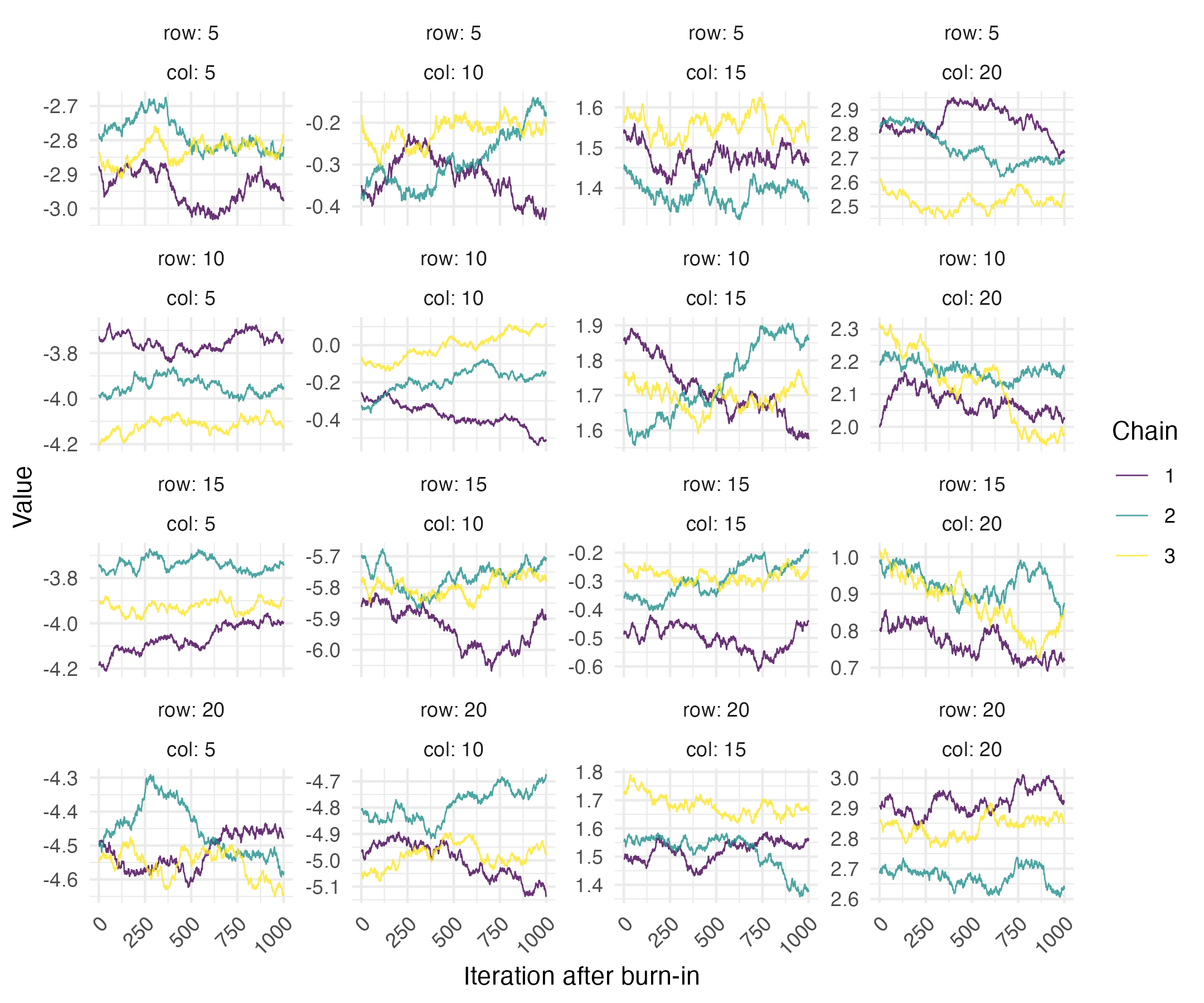}
    \caption{Trace plots of $B$ at select grid locations.}
    \label{fig:trace_1b}
\end{figure}

\begin{figure}[htbp]
    \centering
    \includegraphics[width=0.95\textwidth]{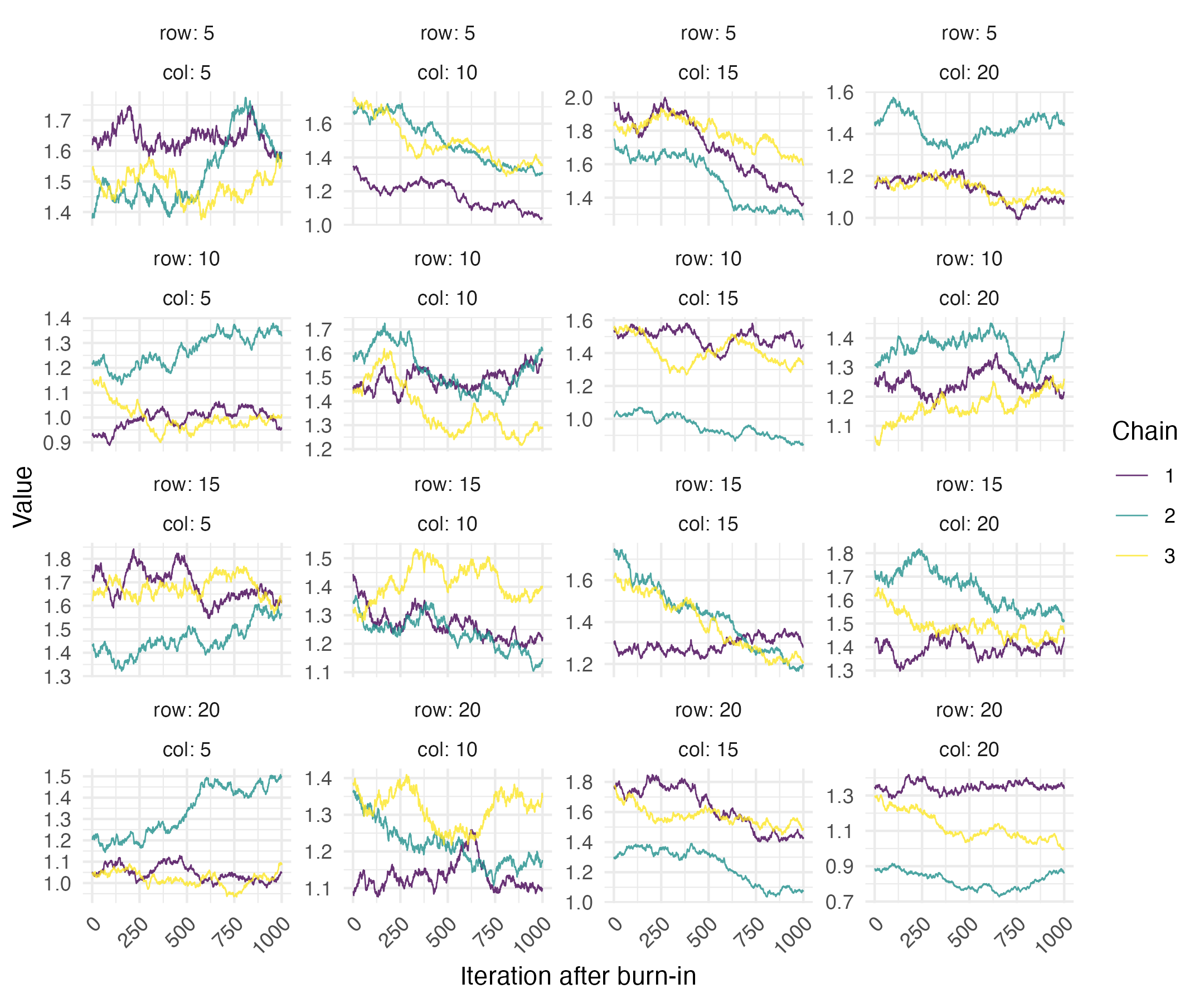}
    \caption{Trace plots of $k$ at select grid locations.}
    \label{fig:trace_1c}
\end{figure}

\begin{figure}[htbp]
    \centering
    \includegraphics[width=0.95\textwidth]{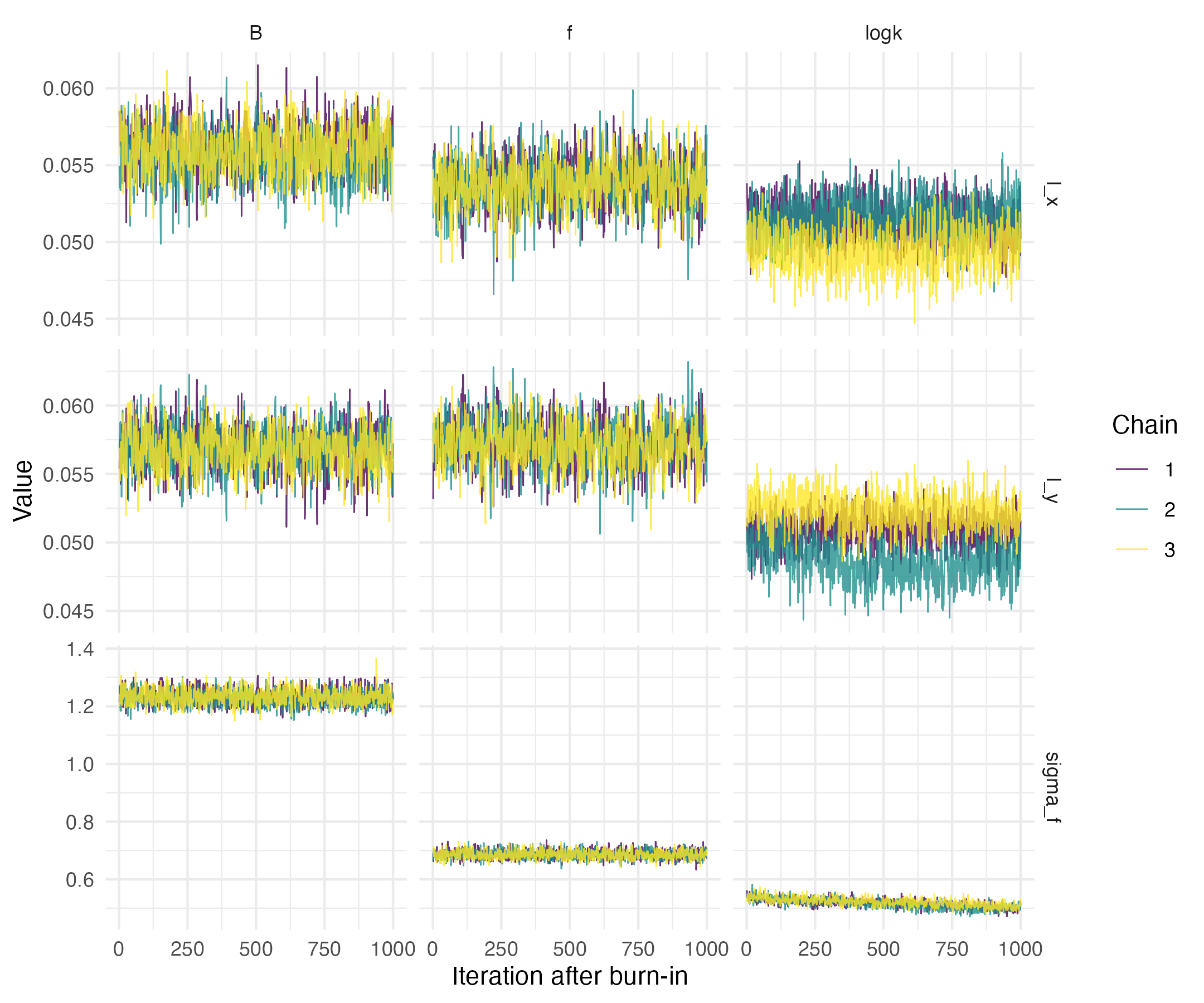}
    \caption{Trace plots of the GP hyperparameters associated with $f$, $B$, and $k$.}
    \label{fig:trace_2}
\end{figure}

\begin{figure}[htbp]
    \centering
    \includegraphics[width=0.7\textwidth]{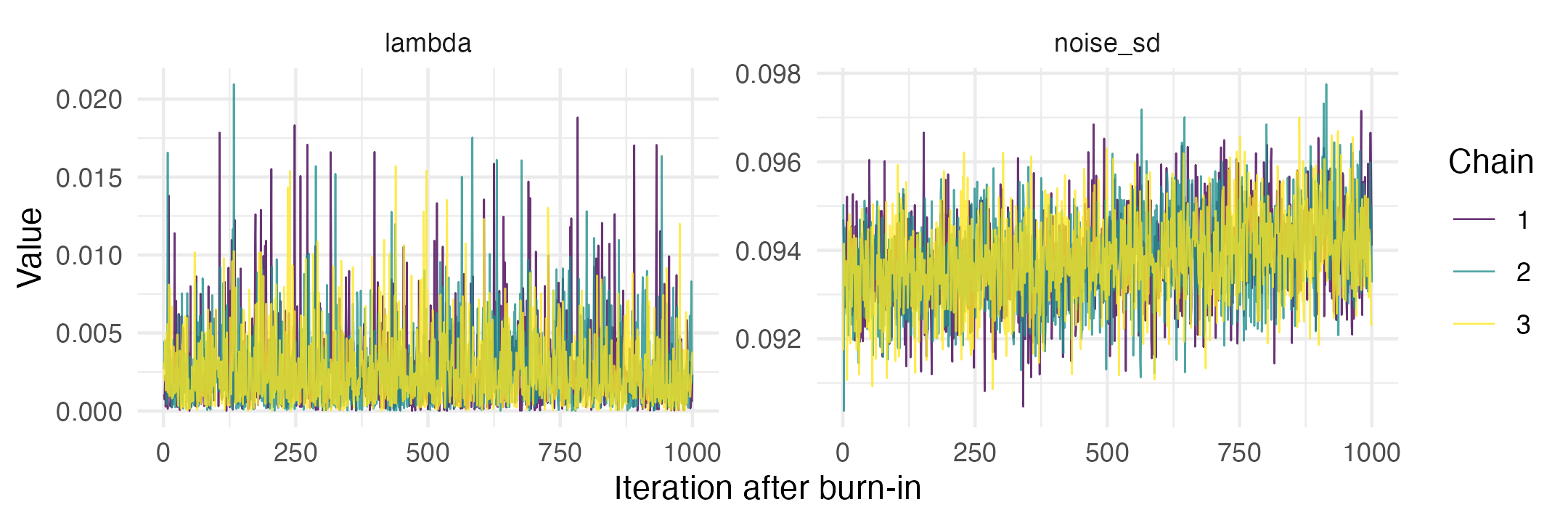}
    \caption{Trace plots of $\sigma$ and $\lambda_k$.}
    \label{fig:trace_3}
\end{figure}

The posterior convergence and mixing is adequate. We show posterior draws for the "middle information" example simulation described in the main paper, in which there were 40,000 total samples drawn with the first 20,000 discarded as burn-in and every 10th subsequent draw retained. Figures \ref{fig:trace_0} through \ref{fig:trace_3} show trace plots of predicted $Y(1)$, $f$, $B$, and $k$ draws at select grid locations, their associated GP hyperparameters, and $\sigma$ and $\lambda_k$. Note $f$, $B$, and $k$ themselves can be practically non-identifiable, which is why we see different modes with different runs, but the hyperparameters, noise, and predictions are all stably converged and identified.

\clearpage

\subsection{Alternative Fitting Approach: Regularized Estimation with Spatial and Asymptotic Penalties}

As an alternative to the Bayesian formulation we introduce a flexible framework based on \emph{regularized nonlinear least squares fitting} with \emph{spatial penalties} and perform cross-validation to learn regularization strengths. 

To promote stable and physically plausible extrapolation, we estimate the parameters \( \theta_{ij} \) using a regularized nonlinear least squares approach that enforces spatial coherence and discourages overly slow convergence. Specifically, we assume that neighboring pixels share similar parameter values, and we penalize solutions where the growth rate \( k_{ij} \) becomes vanishingly small, as this leads to unrealistically flat resolution curves.

Let $\theta$ represents the collection of all pixelwise parameters $\theta_{ij}$ across the image/grid. The loss function minimized over the full image is
\begin{align*}
\mathcal{L}(\theta) = \sum_{(i,j)} \Bigg\{ \, &
\underbrace{\sum_{t=1}^T \left[ y_{ij}(r_t) - f(r_t; \theta_{ij}) \right]^2}_{\substack{\text{data} \\ \text{misfit}}} \\
& + \underbrace{\lambda_{\text{neighbor}} \sum_{(k,\ell) \in \mathcal{N}(i,j)} \left[
(f_{\min, ij} - f_{\min, k\ell})^2 + (B_{ij} - B_{k\ell})^2 + (k_{ij} - k_{k\ell})^2
\right]}_{\substack{\text{spatial} \\ \text{smoothness penalty}}} \\
& + \underbrace{\lambda_{\text{slow}} \cdot \frac{1}{k_{ij}^2}}_{\substack{\text{slow} \\ \text{asymptote penalty}}}
\Bigg\}
\end{align*}
where the first term penalizes misfit to the observed resolution-response data, the second enforces spatial smoothness by penalizing differences in parameters across neighbors \( \mathcal{N}(i,j) \), and the third penalizes small growth rates \( k_{ij} \) to avoid extrapolation artifacts. Optimization proceeds by iteratively updating each pixel’s parameters using a Levenberg--Marquardt algorithm with analytical Jacobians, cycling over the image until convergence.

\subsubsection{Regularization Selection via Hold-Next-Out Cross Validation}

To choose appropriate values of \( \lambda_{\text{neighbor}} \) and \( \lambda_{\text{slow}} \), we adopt a \emph{hold-next-out extrapolation approach}. We split the set of simulation resolutions \( R_\text{all} = \{r_1, \ldots, r_T\} \) at a threshold \( r^*\), using only resolutions \( R_\text{fit} = \{r_t : r_t < r^*\} \) to fit the model at each of the $\lambda$ settings. The withheld data at resolutions \( R_\text{holdout} = \{r_t : r_t \geq r^*\} \) is then used to assess out-of-sample predictive performance. Typically we let \( r^* = r_T \) and hold out only the highest available resolution run from $R_\text{fit}$.

We define the \emph{sum of squared residuals (SSR)} on the held-out data as:
\[
\text{SSR}(\lambda_{\text{neighbor}}^k, \lambda_{\text{slow}}^{\ell}) = \sum_{i,j} \left( y_{ij}(R_\text{holdout}) - \hat{f}(R_\text{holdout}; \hat{\theta}_{ij}^{k\ell}) \right)^2,
\]
and perform a $K \times L$ grid search over values of \( (\lambda_{\text{neighbor}}, \lambda_{\text{slow}}) \), where $\hat{\theta}_{ij}^{k\ell}$ denote the best-fitting parameters associated with the $k$-th value of $\lambda_{\text{neighbor}}$ and the $\ell$-th value of $\lambda_{\text{slow}}$. Among the models within 0.1\% of the minimum SSR, we choose the one closest to the diagonal \( \lambda_{\text{neighbor}} = \lambda_{\text{slow}} \), to encourage balance in regularization.

Using the optimal regularization parameters according to the criteria described above, denoted \( (\hat{\lambda}_{\text{neighbor}}, \hat{\lambda}_{\text{slow}}) \), we refit the model using $R_\text{all}$ to get their associated optimal model parameters $\hat{\theta}_{ij}$.

\subsubsection{Uncertainty Quantification}

To estimate uncertainty at extrapolated resolutions of interest, \( R_\text{extrap} = \{r_t: r_t > r_T\} \), we propagate residual variation from the training fit using an ensemble of perturbed data fits. Assuming homoskedastic noise, we generate confidence intervals for the predicted pixel values \( \hat{f}_{ij}(R_\text{extrap}) \) using local variability in the residuals across fits.

\subsubsection{Getting Results}

We could cross-validate using \( R_{\text{fit}} = \{r_t : r_t < r^*\} \), fitting models across an evenly spaced \( 41 \times 41 \) grid of regularization parameters \((\lambda_{\text{neighbor}}, \lambda_{\text{slow}}) \in [0, 0.4]^2\). The cross-validated sum of squared residuals (SSR) is computed on a holdout set \( R_{\text{holdout}} = \{r^*\} \), and the optimal regularization strengths are selected using the 1 standard error (1se) rule. Specifically, among all parameter combinations whose SSR lies within one standard error of the minimum, we choose the pair \((\lambda_{\text{neighbor}}, \lambda_{\text{slow}})\) that maximizes the total regularization strength, \( \lambda_{\text{neighbor}} + \lambda_{\text{slow}} \), favoring models with both more spatial smoothness and a slower asymptote. We can then visualize the selected regularization strengths across varying model configurations.

\end{document}